\documentclass[
aps, pre,
 amsmath,amssymb,
preprint,%
]{revtex4-2}

\usepackage{graphicx}				
\usepackage{multirow}

\newcommand{\ke}{\mathcal{E}}
\newcommand{\nke}{\widetilde{\ke}}

\newcommand{\besselK}{\mathcal{K}}

\newcommand{\Sn}{\Phi}

\newcommand{\rSn}{ \Sn_\text{rms}}
\newcommand{\mSn}{\left< \Sn \right>}
\newcommand{\snw}{\phi}
\newcommand{\fsnw}{\varphi}
\newcommand{\acsnw}{{\rho_\snw}}
\newcommand{\psdsnw}{{\varrho_\snw}}

\newcommand{\mA}{\left< A \right>}
\newcommand{\m}[1]{\left< #1 \right>}
\newcommand{\tm}[1]{\langle #1 \rangle}

\newcommand{\td}{\tau_\text{d}}
\newcommand{\tw}{\m{w}}
\newcommand{\tp}{\m{w}}
\newcommand{\dt}{\triangle_\text{t}}

\newcommand{\abs}[1]{\lvert #1 \rvert}

\newcommand{\four}[2]{\mathcal{F}\left[{#1}\right](#2)}
\newcommand{\fourinv}[2]{\mathcal{F}^{-1}\left[{#1}\right](#2)}
\newcommand{\fourT}[2]{\mathcal{F}_T\left[{#1}\right](#2)}
\newcommand{\psd}[2]{\mathcal{S}_{#1}(#2)}

\newcommand{\rmd}{\text{d}}
\newcommand{\rms}{\text{rms}}
\newcommand{\rmi}{\text{i}}

\renewcommand{\Re}{\text{Re}}

\let\originalleft\left
\let\originalright\right
\renewcommand{\left}{\mathopen{}\mathclose\bgroup\originalleft}
\renewcommand{\right}{\aftergroup\egroup\originalright}

\renewcommand{\Ref}[1]{Ref.~\onlinecite{#1}}
\newcommand{\Refs}[1]{Refs.~\onlinecite{#1}}
\newcommand{\Eqref}[1]{Eq.~\eqref{#1}}
\newcommand{\Eqsref}[1]{Eqs.~\eqref{#1}}
\newcommand{\Figref}[1]{Fig.~\ref{#1}}
\newcommand{\Figsref}[1]{Figs.~\ref{#1}}
\newcommand{\subFigref}[2]{Fig.~\ref{#1}(#2)}
\newcommand{\subFigsref}[2]{Figs.~\ref{#1}(#2)}
\newcommand{\Secref}[1]{Sec.~\ref{#1}}
\newcommand{\Tabref}[1]{Table~\ref{#1}}
\newcommand{\Appref}[1]{App.~\ref{#1}}

\begin{document}

\title{Deviations from spectral Dirac comb due to semiperiodic pulses}

\author{Audun Theodorsen}
\email{audun.theodorsen@uit.no}
 \affiliation{Department of Physics and Technology, UiT The Arctic University of Norway, N-9037 Troms{\o}, Norway}
 
\author{Gregor Decristoforo}
\email{gregor.decristoforo@uit.no}
\affiliation{Department of Physics and Technology, UiT The Arctic University of Norway, N-9037 Troms{\o}, Norway}
 
\author{Odd Erik Garcia}
\email{odd.erik.garcia@uit.no}
\affiliation{Department of Physics and Technology, UiT The Arctic University of Norway, N-9037 Troms{\o}, Norway}

\date{\today}

\begin{abstract}
In the frequency power spectral density, periodic oscillations appear as a Dirac comb at integer multiples of the frequency of the period. In weakly nonlinear systems or systems close to the primary instability threshold, the periodicity may be perturbed, resulting in deviations from the Dirac comb.
We review and discuss a stochastic model of such semiperiodic fluctuations, while also providing several new results which widen the applicability of the model. The fluctuations are described as a superposition of pulses with a fixed shape. Closed form expressions are derived for the frequency power spectral density in the case of periodic pulse arrivals and a random distribution of pulse amplitudes. In general, the spectrum is a Dirac comb located at multiples of the inverse periodicity time and modulated by the pulse spectrum. Deviations from strict periodicity in the arrivals are considered in two ways: either as a random offset to each periodic arrival (jitter) or as independently distributed waiting times between arrivals (renewal). In this contribution, we show that both ways of including deviations from periodicity remove the Dirac comb with remarkable efficiency, leaving mainly the spectrum of the pulse function. Where the jitter process modulates the mass of the higher harmonics, the renewal process leads to spectral broadening. We clarify the effects of random pulse amplitudes on the frequency power spectrum and demonstrate the applicability of normally distributed waiting times to modeling. Contrary to the previous literature, we argue that negative waiting times do not pose problems for the theory, broadening the applicability of the normal approximation. Randomness in the pulse arrival times is investigated by numerical realizations of the process, and the model is used to describe time series of kinetic energy of fluctuating motions in two-dimensional thermal convection.
\end{abstract}

\maketitle

\section{Introduction}
Weakly nonlinear systems, such as a system with an unstable equilibrium and close to the primary instability threshold, are commonly characterized by semiperiodic oscillations, resulting in a frequency power spectral density (PSD) resembling a Dirac comb \cite{ohtomoExponentialCharacteristicsPower1995,maggsGeneralityDeterministicChaos2011,maggsExponentialPowerSpectra2012,doynefarmerChaoticAttractorsInfinitedimensional1982,libchaberTwoparameterStudyRoutes1983,stonePowerSpectraStochastically1990,klingerRouteDriftWave1997,mensourPowerSpectraDynamical1998,safonovMultifractalChaoticAttractors2002}. 
Far from the linear instability threshold the spectral peaks broaden and in many cases an exponential spectrum results \cite{attenChaoticMotionCoulomb1980,
frischIntermittencyNonlinearDynamics1981,
greensideSimpleStochasticModel1982,
broomheadExtractingQualitativeDynamics1986,
brandstaterStrangeAttractorsWeakly1987,
streettNumericalSimulationAppearance1991,
sigetiSurvivalDeterministicDynamics1995,
sigetiExponentialDecayPower1995,
paulPowerLawBehaviorPower2001,
franzkeDynamicalSystemsExplanation2015,
paceExponentialFrequencySpectrum2008,
paceExponentialFrequencySpectrum2008a,
hornungObservationExponentialSpectra2011,
maggsOriginLorentzianPulses2012,
maggsChaoticDensityFluctuations2015,
zhuChaoticEdgeDensity2017,
ohtomoExponentialCharacteristicsPower1995,
maggsGeneralityDeterministicChaos2011,
maggsExponentialPowerSpectra2012,
doynefarmerChaoticAttractorsInfinitedimensional1982,
libchaberTwoparameterStudyRoutes1983,
stonePowerSpectraStochastically1990,
klingerRouteDriftWave1997,
mensourPowerSpectraDynamical1998,
safonovMultifractalChaoticAttractors2002,
groteDynamicsConvectionDynamos2001,
groteRegularChaoticSpherical2000,
busseConvectiveFlowsRapidly2002,
christensenZonalFlowDriven2001,
christensenZonalFlowDriven2002}.

In this contribution, we investigate the effects of semiperiodic oscillations using a stochastic model that describes the fluctuations as a superposition of pulses with fixed shape and duration but an arbitrary distribution of pulse amplitudes and arrival times.
The model is based on the process known as shot noise or filtered Poisson process, in which the uncorrelated pulses have a sharp rise and an exponential decay, a uniform distribution of arrival times, and an exponential waiting time distribution \cite{parzenStochasticProcesses1999,pecseliFluctuationsPhysicalSystems2000,riceMathematicalAnalysisRandom1944,riceMathematicalAnalysisRandom1945,garciaStochasticModellingIntermittent2016}.

The model is easily extended and modified by different choices of amplitude distributions, pulse functions and arrival time distributions, and has been used to describe autocorrelation functions and power spectra of intermittent fluctuations in turbulent fluids and plasmas \cite{garciaIntermittentFluctuationsTCV2015,walkdenInterpretationScrapeoffLayer2017,kubeComparisonMirrorLangmuir2020,decristoforoIntermittentFluctuationsDue2020,decristoforoNumericalTurbulenceSimulations2021,benczeCharacterizationEdgeScrapeoff2019,zuritaStochasticModelingPlasma2022,ahmedStronglyIntermittentFar2023} as well as from low-dimensional chaotic systems \cite{brunsdenPowerSpectraStrange1987,brunsdenPowerSpectraChaotic1989,holmesTurbulenceCoherentStructures2012, franzkeDynamicalSystemsExplanation2015,paceExponentialFrequencySpectrum2008,paceExponentialFrequencySpectrum2008a, hornungObservationExponentialSpectra2011,maggsOriginLorentzianPulses2012,maggsChaoticDensityFluctuations2015,zhuChaoticEdgeDensity2017,ohtomoExponentialCharacteristicsPower1995,maggsGeneralityDeterministicChaos2011,maggsExponentialPowerSpectra2012}.

Several mathematical results on the autocorrelation function and PSD of the model have previously been established for general arrival time distributions \cite{cantoniMeasurementTechniqueCounting1980,lenemanCorrelationFunctionPower1967,beutlerRandomSamplingRandom1966,beutlerSpectralAnalysisImpulse1968,beutlerStatisticsRandomPulse1971,beutlerTheoryStationaryPoint1966,lowenFractalbasedPointProcesses2005,bartlettSpectralAnalysisPoint1963,sunPowerSpectralDensity2014,lukesStatisticalPropertiesSequences1961}.  
This contribution elucidates (1) the effects of the amplitude distribution on the autocorrelation function and the PSD, (2) the difference between sharply peaked but uncorrelated arrivals (a renewal process) and a strictly periodic process with random time offsets, (3) the rate at which spectral peaks disappear as periodicity decays, (4) the Dirac delta contribution to the PSD at zero frequency, and (5) that positive definite waiting times is not a strict requirement of the theory. The results are applied to semiperiodic bursting of the kinetic energy integral of two-dimensional thermal convection.


This contribution is structured as follows. 
In \Secref{sec:fpp}, we present the stochastic model for a superposition of pulses and  the derivation of its autocorrelation and PSD. Deviations from strict periodicity and their effects on the PSD are considered in \Secref{sec:deviations-period}. 
Then, in \Secref{sec:rb-fpp}, the application to a thermal convection system is presented. A discussion of the results and the main conclusions are given in \Secref{sec:conclusion}. A number of appendices present the formulation and implementation of the thermal convection model as well as mathematical derivations and relations used in the main text.

\section{The power spectral density of a sum of pulses}\label{sec:fpp}
In this section, we present expressions for the autocorrelation and PSD of the stochastic process consisting of randomly as well as periodically arriving, superposed pulses. Much of this development is available in the literature, but we believe our discussion of these results and the figures makes the results more accessible. This section relies in particular on \cite{lenemanCorrelationFunctionPower1967,beutlerRandomSamplingRandom1966,beutlerSpectralAnalysisImpulse1968,beutlerStatisticsRandomPulse1971,beutlerTheoryStationaryPoint1966,lukesStatisticalPropertiesSequences1961,lowenFractalbasedPointProcesses2005}, although the notation follows the previous work by the authors \cite{garciaAutocorrelationFunctionFrequency2017,garciaStochasticModellingIntermittent2016}.

\subsection{Definition of the stochastic model}
The stochastic model considered here consists of a superposition of randomly arriving pulses on a domain of time duration $T$,
\begin{equation}\label{eq:snn}
  \Sn(t) = \sum_{k=1}^{K(T)} A_k \snw\left(\frac{t-s_k}{\tau_k}\right).
\end{equation}
All pulses are assumed to have the same functional form $\snw$. The amplitudes $A$ and the pulse duration times $\tau$ are randomly distributed, and $s$ denotes the arrival times according to the stationary point process $K(t)$. We require that the increments $K(t+t')-K(t')$ are stationary and have finite mean and variance \cite{beutlerTheoryStationaryPoint1966}. We denote the waiting times as $w_k = s_{k}-s_{k-1}$, so there is on average a time $\tw$ between the pulses, which gives the \emph{intensity} of the point process as $\tm{K(t+t') - K(t')} = t/\tw$. We further denote the degree of pulse overlap with the \emph{intermittency parameter} $\gamma = \tm{\tau}/\tw$. For $\gamma \ll 1$, pulses are well separated and the signal appears intermittent. For $\gamma \gg 1$, many new pulses arrive during the average lifetime of a pulse, leading to a signal which appears more normally distributed.

\subsubsection{The sum of waiting times}
We use $S_k$ to denote the sum of $k$ waiting times:
\begin{equation}
  S_k = s_{1+k} - s_1 = \sum\limits_{m=2}^{1+k} w_m \overset{\text{dist.}}{=} \sum\limits_{m=n+1}^{n+k} w_m = s_{n+k}-s_n\, \forall\, 1\leq n\leq K(T)-k.
  \label{eq:def-sk}
\end{equation}
where $\overset{\text{dist.}}{=}$ indicates equality in distribution, which holds for all starting arrivals $s_n$ due to the stationarity of the process. We denote the distribution of $S_k$ by $p_{S}(\sigma;k)$, defined as the derivative of the probability that $S_k \leq \sigma$:
\begin{equation}
  p_S(\sigma;k) = \frac{\partial}{\partial \sigma} P[S_k \leq \sigma].
  \label{eq:def-Sk-pdf}
\end{equation}
Its characteristic function is denoted by $\psi_S(\nu;k)=\m{\exp(\rmi S_k \nu)}$. We have that
\begin{align}
  p_S(\sigma;0)=\delta(\sigma), &\quad \psi_S(\nu;0)=1, \label{eq:S0-convention} \\
  p_S(\sigma;1)=p_w(\sigma), &\quad \psi_S(\nu;1)=\psi_w(\nu), \label{eq:S1-convention}\\ \label{eq:S-convention}
  p_S(\sigma;-k)=p_S(-\sigma;k), & \quad \psi_S(\nu;-k)=\psi_S(-\nu;k)=\psi_S^*(\nu;k).
\end{align}
Here and in the following, $p_w$ and $\psi_w$ denote the PDF and characteristic function of the waiting times, respectively, and we denote complex conjugation by an asterisk. The first line, \Eqref{eq:S0-convention}, follows from $S_0 = 0$ and the last, \Eqref{eq:S-convention}, from
\begin{equation}
   S_{-k} = s_{n-k} - s_n = -\left(s_n - s_{n-k}\right)\overset{\text{dist.}}{=} -\left(s_{n+k}-s_n\right) = - S_k.
  \label{eq:Sk-negative}
\end{equation}
where we choose $n$ such that $n>k$.

\subsubsection{The pulse function}
We normalize the pulse function such that
\begin{equation}
    \int_{-\infty}^\infty \rmd \theta \,\abs{\snw(\theta)} = 1.
\end{equation}
where $\theta$ is a unitless variable. Using the Fourier transform $\mathcal{F}$ defined in \Appref{app:def_four} and the unitless variable $\vartheta$, we also introduce the notation
\begin{equation}
  \fsnw(\vartheta) = \four{\snw}{\vartheta},
  \label{eq:four_pulse}
\end{equation}
\begin{equation}\label{eq:corrf_pulse}
\acsnw(\theta) = \frac{1}{I_2} \int\limits_{-\infty}^{\infty} \rmd u\,\snw(u) \snw(u+\theta)
\end{equation}
and 
\begin{equation}\label{eq:spectrum_pulse}
\psdsnw(\vartheta) = \frac{1}{I_2} \abs{\fsnw(\vartheta)}^2
\end{equation}
for the Fourier transform, autocorrelation and PSD of the pulse function, respectively. Here, the pulse integral of order $n$ is defined by
\begin{equation}
    I_n = \int_{-\infty}^\infty \rmd \theta\,[\snw(\theta)]^n .
\end{equation}
Note that the functions $\acsnw$ and $\psdsnw$ form a Fourier transform pair. We further note that $\acsnw(0) = 1$ and $\psdsnw(0) = I_1^2/I_2$. 

\subsection{The finite-time power spectral density}\label{app:psd-finite-time}

To obtain the PSD, we start from \Eqref{eq:snn}, and take the finite-time Fourier transform as defined in \Appref{app:def_four},
\begin{equation}
  \fourT{\Sn}{\omega} = \int\limits_{0}^{T} \rmd t\, \exp(-i \omega t) \Sn = \sum\limits_{k=1}^{K(T)} A_k \int\limits_{0}^{T} \rmd t\, \exp(-i \omega t)  \snw\left(\frac{t-s_k}{\tau_k}\right),
  \label{eq:fourier_fpp_start}    
\end{equation}
where $\omega = 2\pi f$ is the angular frequency. Changing integration variables to $u(t) = (t-s_k)/\tau_k$ and ignoring end effects in the integral by assuming $T/\tau_k \gg 1$, we get 
\begin{equation}
  \fourT{\Sn}{\omega} = \sum\limits_{k=1}^{K(T)} A_k \tau_k \fsnw(\tau_k \omega) \exp(-i \omega s_k).
  \label{eq:fourier_fpp}    
\end{equation}
The PSD is then given by
\begin{multline}
  \psd{\Sn}{\omega} = \lim_{T \to \infty} \frac{1}{T} \m{\abs{\fourT{\Sn}{\omega}}^2} \\= \lim_{T \to \infty}\frac{1}{T} \m{ \sum\limits_{k,l=1}^{K(T)} A_k A_l \tau_k \tau_l \fsnw(\tau_k \omega)\fsnw^*(\tau_k \omega) \exp(i \omega (s_l -s_k))}
  \label{eq:PSD_fpp_start}
\end{multline}
where the angular brackets denote averaging over all random variables. Assuming the arrivals are independent of the amplitudes and pulse durations (but not necessarily that amplitudes and pulse durations are independent), and using that $s_l - s_k = S_{l-k}$, we get
\begin{equation}
  \psd{\Sn}{\omega} = \lim_{T \to \infty} \frac{1}{T} \sum_{K=0}^{\infty} P_K(K;T)\sum\limits_{k,l=1}^{K}\m{ A_k A_l \tau_k \tau_l \fsnw(\tau_k \omega)\fsnw^*(\tau_l \omega)} \psi_S(\omega;l-k).
  \label{eq:PSD_fpp}
\end{equation}
Here, $P_K(K;T)$ is the probability mass function of $K(T)$ where we retain the explicit dependence on the parameter $T$. We now assume that the amplitudes and pulse durations are pairwise uncorrelated and independent of each other so that $\m{ A_k A_l \tau_k \tau_l \fsnw(\tau_k \omega)\fsnw^*(\tau_l \omega)} = I_2 \m{A^2 \tau^2 \psdsnw(\tau \omega)}$ if $k=l$ and $\abs{\m{A \tau \fsnw(\tau \omega)}}^2$ if $k\neq l$. If the amplitudes and pulse durations are independent, these expressions read $I_2 \m{A^2} \m{\tau^2 \psdsnw(\tau \omega)}$ if $k=l$ and $\mA^2 \abs{\m{\tau \fsnw(\tau \omega)}}^2$ if $k\neq l$.
 In the double sum in \Eqref{eq:PSD_fpp} there are $K$ terms for which $k=l$ and $K(K-1)$ terms for which $k \neq l$. Summing over all these terms, we have that 
\begin{multline}
  \psd{\Sn}{\omega} = \frac{1}{\tw} I_2 \m{ A^2 \tau^2 \psdsnw(\tau \omega)}\\+ \abs{\m{A \tau \fsnw(\tau \omega)}}^2 \lim_{T \to \infty} \frac{1}{T} \sum_{K=0}^{\infty} P_K(K;T)\sum\limits_{\substack{k,l=1\\k\neq l}}^{K}\psi_S(\omega;l-k).
  \label{eq:PSD_fpp_2}
\end{multline}
Using the the property of the characteristic function in \Eqref{eq:S-convention} and manipulating the sums, it may be seen that
\begin{multline}
  \psd{\Sn}{\omega} = \frac{1}{\tw} I_2 \m{ A^2 \tau^2 \psdsnw(\tau \omega)}\\+ \abs{\m{A \tau \fsnw(\tau \omega)}}^2 \lim_{T \to \infty} \frac{1}{T} \sum_{K=0}^{\infty} P_K(K;T)\sum_{k=1}^K 2 (K-k)\text{Re}\left[\psi_S(\omega;k)\right].
\end{multline}
where the operator $\text{Re}[\cdot]$ denotes the real part of the argument. Thus the PSD is a real valued function, as required. This equation should be compared to Eq.~(2.10) or (2.11) in \Ref{beutlerStatisticsRandomPulse1971}. 

We assume that for large $T$, the distribution $P_K$ becomes very narrow and $K$ may be replaced by $\m{K}$. As an example where this assumption holds, for waiting times according to a renewal process with finite mean value and variance, $K(T) \sim \mathcal{N}(T/\tw, w_\rms^2 T / \tw^3)$ for large $T$ \cite{parzenStochasticProcesses1999}. Thus for large $T$ $\m{K} \gg K_\rms$, implying that $P_K$ is indeed a narrow distribution around $\m{K}$. This gives
\begin{equation}
  \psd{\Sn}{\omega} = \frac{1}{\tw} I_2 \m{ A^2 \tau^2 \psdsnw(\tau \omega)}\\+ \abs{\m{A \tau \fsnw(\tau \omega)}}^2 \lim_{T \to \infty} \frac{1}{T} \sum_{k=1}^{\m{K}} 2 (\m{K}-k)\text{Re}\left[\psi_S(\omega;k)\right].
  \label{eq:PSD_fpp_finite_T}
\end{equation}
The authors of \Ref{beutlerStatisticsRandomPulse1971} argue that for $T \to \infty$, the $k/T$-term disappears and we get the equivalent of the following equation \cite{beutlerStatisticsRandomPulse1971,lenemanCorrelationFunctionPower1967}
\begin{equation}
  \psd{\Sn}{\omega} = \frac{1}{\tw} I_2 \m{ A^2 \tau^2 \psdsnw(\tau \omega)}\\+ \frac{2}{\tw}\abs{\m{A \tau \fsnw(\tau \omega)}}^2\sum_{k=1}^\infty\text{Re}\left[\psi_S(\omega;k)\right], \quad \omega \neq 0.
  \label{eq:PSD_fpp_inf}
\end{equation}
where we claim this is only valid for $\omega \neq 0$ since the last term in \Eqref{eq:PSD_fpp_inf} diverges for $\omega = 0$. This is due to the long-time behavior of the correlation function, $\lim_{\tau \to \infty} \m{\Phi(t) \Phi(t+\tau)} = \mSn^2 = \m{A \tau}^2 I_1^2/\tw^2$, so the PSD which is the Fourier transform of the correlation function should contain a term $2 \pi \mSn^2 \delta(\omega)$. In \Ref{lowenFractalbasedPointProcesses2005} this term was added \emph{ad hoc} for renewal processes after employing (the equivalent of) \Eqref{eq:PSD_fpp_inf}. It is, however, not clear to the authors that this term in general may be extracted from the infinite sum, nor that it is simply absent and may be added in later. Indeed, as we will see below, the $\delta(\omega)$-contribution would simply be missing for the Poisson process, but is present for the periodic pulse train if using \Eqref{eq:PSD_fpp_inf}. Therefore, we base our further development on \Eqref{eq:PSD_fpp_finite_T} instead.

\subsubsection{Degenerate duration time distribution}\label{sec:degenerate_tau}

In the analysis of thermal convection in \Secref{sec:rb-fpp}, degenerate duration times $p_\tau(\tau) = \delta(\tau-\td)$ will be assumed. This simplifies the expressions above. Focusing on \Eqref{eq:PSD_fpp_finite_T}, we have
\begin{equation}
  \psd{\Sn}{\omega} = \gamma \td I_2 \m{ A^2} \psdsnw(\td \omega)+ \mA^2 \td^2 I_2 \psdsnw(\td \omega) \lim_{T \to \infty} \frac{1}{T} \sum_{k=1}^{\m{K}} 2 (\m{K}-k)\text{Re}\left[\psi_S(\omega;k)\right].
  \label{eq:PSD_fpp_finite_T_deg_tau}
\end{equation}
which may be seen as the PSD of the pulse function, $\td I_2 \psdsnw(\td \omega)$, multiplied by the PSD of a train of delta pulses located at the arrival times with mass according to the amplitudes,
\begin{align}
  \psd{\Sn}{\omega} &= I_2 \psdsnw(\td \omega) \lim_{T\to \infty} \frac{1}{T} \m{\abs{\fourT{F}{\omega}}^2}, \\
  F(t) &= \sum_{k=1}^{K(T)} A_k \delta\left(\frac{t-s_k}{\td}\right).
  \label{eq:S_split}
\end{align}
This is consistent with the observation that for a degenerate distribution of durations, the stochastic process is a convolution between the pulse function and the delta pulse train $F$. 

This simplification also allows for a reformulation of the contribution of the amplitudes using the rms-value, giving
\begin{equation}
  \psd{\Sn}{\omega} = \gamma \td I_2 A_\rms^2 \psdsnw(\td \omega)+ \mA^2 \td^2 I_2 \psdsnw(\td \omega) \lim_{T \to \infty} \frac{1}{T} \sum_{k=1}^{\m{K}} 2 (\m{K}-k+1)\text{Re}\left[\psi_S(\omega;k)\right].
  \label{eq:PSD_fpp_finite_T_deg_tau_rms}
\end{equation}
where we have used $1/\m{w} = \lim_{T\to\infty} \m{K}/T = \lim_{T\to\infty}\sum_{k=1}^{\m{K}}T^{-1}$. There is a conceptual advantage in using $A_\rms$ instead of $\m{A^2}$. While the latter is positive for any distribution of $A$, $A_\rms$ disappears for a degenerate distribution of amplitudes, leaving only the second term in \Eqref{eq:PSD_fpp_finite_T_deg_tau_rms}. On the other hand, the influence of the arrival time distribution disappears for $\mA=0$. This fact in the case of a renewal process with $A = \pm 1$ with equal probability was discussed in \Ref{brunsdenPowerSpectraChaotic1989}, but the result holds for any arrival time distribution and any amplitude distribution with $\mA=0$, as long as the duration times are degenerate or even just independent of $A$, as may be seen from \Eqref{eq:PSD_fpp_finite_T}. 

\subsection{The power spectral density for renewal processes}\label{sec:renewal}
A renewal process is a point process $K$ with independent and identically distributed waiting times $w$, but where the distribution of $w$ may differ from the exponentially distributed waiting times of the Poisson process. The case of renewal arrivals has been considered in for example \Refs{lukesStatisticalPropertiesSequences1961,lowenFractalbasedPointProcesses2005,cantoniMeasurementTechniqueCounting1980,brunsdenPowerSpectraStrange1987,brunsdenPowerSpectraChaotic1989,holmesTurbulenceCoherentStructures2012}. 

Here we show that the $\delta(\omega)$-contribution is kept by not taking the limit before calculating the sums for both renewal processes and the periodic process. Assuming renewal or periodic arrivals, we have $\psi_S(\omega;k) = \psi_w(\omega)^k$ for $k > 0$ and the last term in \Eqref{eq:PSD_fpp_finite_T} contains
\begin{equation}
  \lim_{T\to\infty}\frac{1}{T} \sum_{k=1}^{\m{K}} 2 (\m{K}-k)\text{Re}\left[\psi_w(\omega)^k\right]
  =\lim_{T\to\infty}2 \Re\left[\frac{\psi_w}{\left(1-\psi_w \right)^2}\frac{\psi_w^{\m{K}}-1}{T} + \frac{1}{\tw} \frac{\psi_w}{1-\psi_w} \right] .
  \label{eq:FPP_start_renewal}
\end{equation}
This expression should be compared to Eq.~(21) in \Ref{lukesStatisticalPropertiesSequences1961}. Since the central limit theorem gives 
\begin{equation}
    \psi_w(\omega)^{\m{K}} = \psi_S(\omega;\m{K}) \approx \exp\left(\rmi \omega \tw \m{K} - \omega^2 w_\rms^2 \m{K}\right),
\end{equation}
 the first term in \Eqref{eq:FPP_start_renewal} does not survive in the limit $T \to \infty$ and so only the second is kept \cite{lukesStatisticalPropertiesSequences1961,lowenFractalbasedPointProcesses2005}. However, this is only valid for $\omega \neq 0$, since $\psi_w(0) = 1$ may cause the first term to diverge.
 
 Letting $\omega \ll 1$,  we have to first order that $\psi_w(\omega) \approx 1 + \rmi \tw \omega$ and we approximate $\psi_w(\omega)^{\m{K}} \approx \exp\left(\rmi \omega \tw \m{K}\right)=\exp\left(\rmi \omega T\right)$. The first term is then
\begin{align*}
  &\lim_{T\to\infty}2 \Re\left[\frac{\psi_w}{\left(1-\psi_w \right)^2}\frac{\psi_w^{\m{K}}-1}{T}  \right] \\
  \approx&\lim_{T\to\infty}2 \Re\left[\frac{1}{-\tw^2 \omega^2}\frac{e^{iT\omega}-1}{T}\right] \\
  =&\lim_{T\to\infty}\frac{2}{\tw^2} \frac{1-\cos(\omega T)}{T\omega^2} \\
  =&\frac{2\pi}{\tw^2} \delta(\omega) 
\end{align*}
where the last equality is from \Appref{app:delta-func}. Thus for the renewal process, \Eqref{eq:PSD_fpp_finite_T} gives
\begin{multline}
  \psd{\Sn}{\omega} = \frac{2\pi}{\tw^2} \m{A \tau}^2 I_1^2\delta(\omega)+\frac{1}{\tw} I_2 \m{ A^2 \tau^2 \psdsnw(\tau \omega)}+ \\ \frac{2}{\tw}\abs{\m{A \tau \fsnw(\tau \omega)}}^2 \text{Re}\left[\frac{\psi_w(\omega)}{1-\psi_w(\omega)}\right]
  \label{eq:PSD_fpp_renewal}
\end{multline}
since $\four{\snw}{0}=I_1$. This expression is in agreement with \Ref{lowenFractalbasedPointProcesses2005}, and it is straightforward to show that the first term is $2\pi \mSn^2 \delta(\omega)$.

Concerning large $\omega$, many common probability distributions have characteristic functions which go to zero in absolute value for large argument. If this is the case, the fraction in the last term of \Eqref{eq:PSD_fpp_renewal} approaches zero and only the middle term survives, giving a spectrum which does not depend on the waiting time distribution. Degenerately distributed waiting times, corresponding to periodic pulse arrivals, do not have characteristic functions which decay to zero, so this result does not hold for these as will be seen in \Secref{sec:periodic-arrivals}. 

\subsubsection{Degenerate duration time distribution}
For degenerately distributed durations, we may for reference write the PSD in a slightly different form which highlights the contribution of $A_\rms$,
\begin{align}
  \psd{\Sn}{\omega} &= 2\pi\gamma^2 \m{A}^2 I_1^2\delta(\omega)+\td \gamma A_\rms^2 I_2 \psdsnw(\td \omega) + \td \gamma \mA^2 I_2 \psdsnw(\td \omega) \text{Re}\left[\frac{1+\psi_w(\omega)}{1-\psi_w(\omega)}\right]. 
  \label{eq:psd-snn-inf-renewal-arms}
\end{align}

\subsubsection{Polar form}
For some waiting time distributions (for example the family of stable distributions \cite{nolanUnivariateStableDistributions2020}), the characteristic function may be easily written in polar form $\psi_w(\omega) = r(\omega) e^{\rmi \theta(\omega)}$. Then \eqref{eq:PSD_fpp_renewal} may be written as
\begin{multline}
  \psd{\Sn}{\omega} = \frac{2\pi}{\tw^2} \m{A \tau}^2 I_1^2\delta(\omega)+\frac{1}{\tw} I_2 \m{ A^2 \tau^2 \psdsnw(\tau \omega)}+ \\ \frac{2}{\tw}\abs{\m{A \tau \fsnw(\tau \omega)}}^2\frac{r \cos(\theta)-r^2}{1+r^2 - 2 r \cos(\theta)}. 
  \label{eq:PSD_fpp_renewal_polar}
\end{multline}

\subsubsection{The Poisson process}\label{sec:poisson}
For the Poisson process, the waiting times are well known to be exponentially distributed, $\psi_w(\omega) = (1-\rmi \omega \tw)^{-1}$ \cite{parzenStochasticProcesses1999}. In this case the term inside $\Re[\cdot]$ in \Eqref{eq:PSD_fpp_renewal} is purely imaginary and we recover the well-known result
\begin{equation}
  \psd{\Sn}{\omega} = \frac{2\pi}{\tw^2} \m{A \tau}^2 I_1^2\delta(\omega)+\frac{1}{\tw} I_2 \m{ A^2 \tau^2 \psdsnw(\tau \omega)}
  \label{eq:PSD_fpp_poisson}
\end{equation}
A Poisson point process gives a flat spectrum, so the only frequency variation for the stochastic process is due to the pulse function. Here, it was necessary to include the correct $T\to\infty$-limit to recover the contribution of the mean value. The effect of the distribution of $\tau$ was discussed extensively in \Refs{garciaPowerLawSpectra2017,garciaAutocorrelationFunctionFrequency2017}.

\subsubsection{Gamma distributed waiting times}\label{sec:gamma-wait}
\begin{figure}
\includegraphics[width=\textwidth]{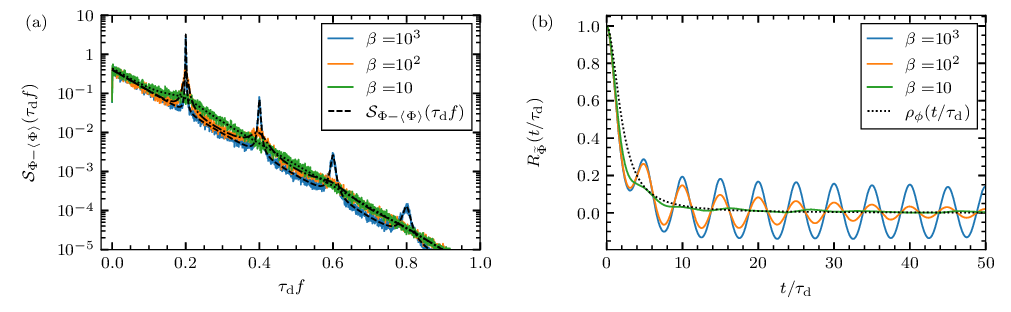}
\caption{Power spectral density (left) and autocorrelation function (right) for a sum of Lorentzian pulses with Gamma distributed waiting times and exponentially distributed amplitudes for the case $\tw = 5 \td$. The analytical expressions are represented by the black lines. For the numerical realizations, $T = 10^5\td$ and the sampling time is $10^{-2}\td$.}\label{fig:psd_ac_gamma}
\end{figure}

A generalized form of the Poisson arrivals are Gamma distributed waiting times with mean value $\tw$ and shape parameter $\beta$, giving
\begin{align}
  \psi_w (\omega) &= {(1-\rmi\omega\tw/\beta)}^{-\beta} , \\
  \psi_S (\omega;k) &= {(1-\rmi\omega\tw/\beta)}^{-k\beta}={(1-\rmi\omega k\tw/ k\beta)}^{-k\beta} . 
\end{align}
That is, $S_k$ is also Gamma distributed, with mean value $k \tw$ and shape parameter $k\beta$. For fixed $\tw$ and $\beta \to \infty$, we recover the periodic case with period $\tw$ since $\psi_w(\omega) \to \exp(-\rmi\omega\tw)$. 
For $\beta = 1$, the waiting times are exponentially distributed. In \Figref{fig:psd_ac_gamma} we present the PSD in \Eqref{eq:PSD_fpp_poisson} for Gamma distributed waiting times with various values of the shape parameter $\beta$. The analytic expressions agree with the numerical results. The spectral peaks vanish for remarkably large $\beta$ (corresponding to remarkably narrow distributions): for $\beta=10$ there is only a hint of the first spectral peak, while for $\beta=10^2$, only two peaks are clearly visible. For $\beta=10^3$, many spectral peaks are visible although they display significant broadening. As will be seen in \Secref{sec:deviations-period}, $\m{w}/w_\rms\geq 10^2$, corresponding to $\beta \geq 10^4$, is required to have an approximately periodic process as seen from the power spectral density.

\subsection{Periodic arrivals}\label{sec:periodic-arrivals}
For periodic arrivals, $\psi_w(\omega) = \exp(\rmi \omega \tw)$, and the arguments leading to \Eqref{eq:PSD_fpp_renewal} in \Secref{sec:renewal} regarding $\omega \ll 1$ hold for all $\omega - 2 \pi n/\tw \ll 1$ where $n$ is an integer. Further, it is straightforward to show that the second term inside $\Re[\cdot]$ is identically equal to $-1/2$. Therefore we get from \Eqref{eq:PSD_fpp_finite_T}
\begin{multline}
  \psd{\Sn}{\omega} = \frac{1}{\tw} I_2 \m{ A^2 \tau^2 \psdsnw(\tau \omega)}- \frac{1}{\tw}\abs{\m{A \tau \fsnw(\tau \omega)}}^2\\+ 
\frac{2\pi}{\tw^2}\abs{\m{A \tau \fsnw(\tau \omega)}}^2 \sum_{n=-\infty}^{\infty} \delta(\omega - 2 \pi n/\tw)
  \label{eq:PSD_fpp_finite_T_periodic}
\end{multline}
This result may also be obtained from \Eqref{eq:PSD_fpp_inf} by using the relation $\sum_{k=-\infty}^\infty \exp(2\pi\rmi k x) = \sum_{k=-\infty}^\infty \delta(x-k)$ \cite{richardsTheoryDistributionsNontechnical1990}. 

\subsubsection{The effect of the amplitude distribution}
\begin{figure}
\includegraphics[width=\textwidth]{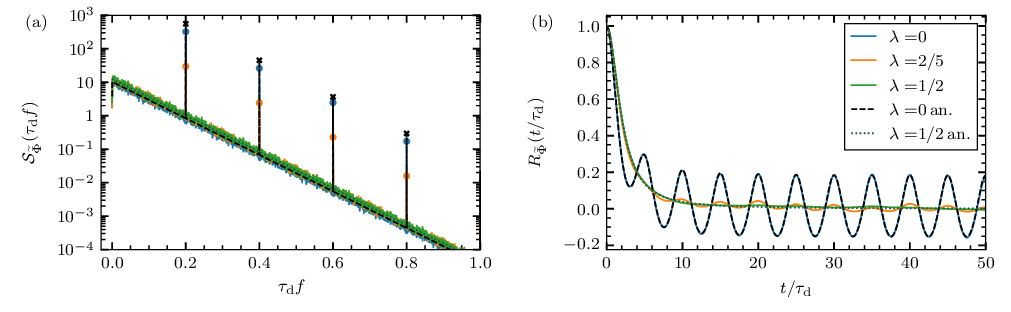}
\caption{Power spectral density (left) and autocorrelation function (right) for a sum of Lorentzian pulses with periodic arrival times and Laplace distributed amplitudes with various asymmetry parameters $\lambda$ in the case $\tp = 5 \td$ and $A_\rms=1$. The analytical expressions are given by the black and gray dashed lines, respectively. The filled circles indicate the peak of the numerical PSD, while the black crosses indicate the mass of the delta function. For the numerical realizations, $T = 10^5\td$ and the sampling time is $10^{-2}\td$.}\label{fig:different_asym}
\end{figure}

In \Figref{fig:different_asym}, we present the effect of the amplitude distribution on the empirical PSD (left) and autocorrelation function (right) of a process with asymetrically Laplace distributed amplitudes, see \Appref{app:alap-dist}. Asymmetry parameter $\lambda = 0$ corresponds to exponentially distributed amplitudes while $\lambda = 1/2$ corresponds to symmetrically distributed Laplace amplitudes with vanishing mean. The analytical expression given by \Eqref{eq:PSD_fpp_finite_T_periodic} and \Eqref{eq:ac-per-deg-td} for the two cases $\lambda = 0$ and $\lambda = 1/2$ are presented by the black and grey dashed lines, respectively. On the left-hand side of \Figref{fig:different_asym}, the Dirac comb is seen in the cases with $\lambda \neq 1/2$ distributed amplitudes, but is canceled out in the case of pulse amplitudes with vanishing mean. As $\lambda$ goes from $0$ to $1/2$, the mass of the Dirac comb gradually decays. The same effect is seen in the autocorrelation function, shown on the right-hand side of \Figref{fig:different_asym}.


\subsubsection{Jittered periodic arrivals}
We now consider periodic arrivals with jitter, where the arrival times are perturbed relative to their original position \cite{beutlerRandomSamplingRandom1966}.
In this case, $s_k \to s_k + X_k$ where the $X$'s are independent and identically distributed random variables. Then $S_k = s_{n+k}-s_{n} \to s_{n+k}+X_{n+k}-s_n-X_n = k \tp + X_1 - X_0$ where the equality holds in distribution, and we have
\begin{align}
  \psi_w (\omega) &= \exp(\rmi \omega \tp) \abs{\psi_X(\omega)}^2 \\
  \psi_S (\omega;k) &= \exp(k \rmi \omega \tp)\abs{\psi_X(\omega)}^2 \\
  \td p_S (\td u;k) &= \tp \left[ p_X * p_{-X} \right](\td u - \tp k)
\end{align}
where $\psi_X$ is the characteristic function of $X$. Note that this expression does not fulfill the convention $\psi_S(\omega,0)=1$. Referring back to \Eqref{eq:PSD_fpp_2}, we see that the jitter process just gives a multiplicative factor to the last term, since $k=l$ has been taken out of the equation.   Thus we straightforwardly get
\begin{multline}
  \psd{\Sn}{\omega} = \frac{1}{\tw} I_2 \m{ A^2 \tau^2 \psdsnw(\tau \omega)}- \frac{1}{\tw}\abs{\m{A \tau \fsnw(\tau \omega)}\psi_X(\omega)}^2\\+ 
\frac{2\pi}{\tw^2}\abs{\m{A \tau \fsnw(\tau \omega)}\psi_X(\omega)}^2 \sum_{n=-\infty}^{\infty} \delta(\omega - 2 \pi n/\tw).
  \label{eq:PSD_fpp_jitter}
\end{multline}
If $p_X$ admits the delta distribution as a limiting case, we recover purely periodic arrivals, while for sufficiently broad jitter distributions, there is some probability that consecutive arrivals will be reordered, with $s_{k+1}<s_k$.  We emphasize that this is still a very restrictive formulation: each arrival is guaranteed to be centered on the time corresponding to the periodic arrival time, and the number of arrivals in a given interval is fixed up to end effects. We will show in \Secref{sec:deviations-period} that at least for normally distributed jitter with $w_\rms \leq 5\tw$, reordering of arrivals has no effect on the PSD. We also note that degenerate arrival times and $\mA=0$ will remove the effect of the jitter, just as it removes the effect of the arrival time distribution.
\subsection{Deviations from periodicity}\label{sec:deviations-period}
In this section, we will consider deviations from strict periodicity in two different ways, using a renewal process and jittered periodic arrivals. It will be demonstrated that the Dirac comb is present as long as periodic arrivals are maintained. However, slight deviations from periodic arrivals efficiently remove most higher harmonics in the Dirac comb. We will also demonstrate that a normal distribution of waiting times is a good approximation to a broad selection of waiting time distributions.

\subsubsection{Comparison between renewal waiting times and jittered periodic arrivals}\label{sec:renewal_vs_jitter}
As an instructive example on the difference between periodic arrivals with jitter and a nearly periodic renewal process, we consider a normal distribution around the periodicity in both cases. Normally distributed inter-event times were considered in \Refs{brunsdenPowerSpectraChaotic1989,brunsdenPowerSpectraStrange1987}. We use degenerately distributed durations, exponentially distributed amplitudes and a Lorentzian pulse function.

\begin{figure}
\includegraphics[width=\textwidth]{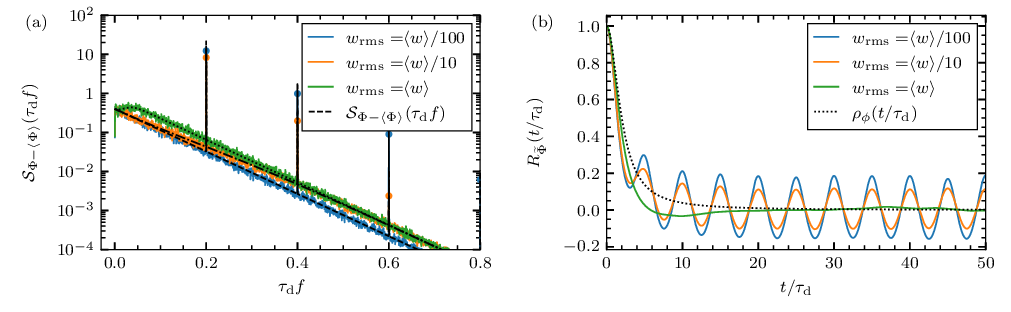}
  \caption{Power spectral density (left) and autocorrelation function (right) for a sum of Lorentzian pulses with periodic arrival times with a period of  $\tw = 5 \td$ and normal jitter, and various values of $w_\rms$. Left, the analytical expression in \Eqref{eq:jitter-gauss} is presented by the black dashed line. Right, the black dotted line denotes the pulse autocorrelation. For the numerical realizations, $T = 10^5\td$ and the sampling time is $10^{-2}\td$.} \label{fig:jitter-gauss}
\end{figure}

If we have periodic arrivals with normally distributed jitters, we get $\abs{\psi_X(\omega)}^2 = \exp(- w_\rms^2 \omega^2)$, so the expression in \Eqref{eq:PSD_fpp_jitter} becomes
\begin{align}
  \psd{\Sn}{\omega} &= \td \gamma I_2 \psdsnw(\td\omega) \left[\m{A^2} - \mA^2 \exp(- w_\rms^2 \omega^2)\right] \nonumber \\
  &+ 2 \pi \gamma^2 \m{A}^2 I_2 \psdsnw(\td\omega)\exp(- w_\rms^2 \omega^2)\sum\limits_{n=-\infty}^\infty \delta(\omega - 2\pi n/\tw).
  \label{eq:jitter-gauss}
\end{align}
The normal distribution modifies the pulse spectrum in addition to modulating the Dirac comb. The result is presented in \Figref{fig:jitter-gauss}. Here, both the effect of the modulation of the delta peaks as well as the modulation of the first term are clearly seen. For $w_\rms = \tw / 100$, the modulation is very close to one and the amplitude factor in the first term of \Eqref{eq:jitter-gauss} is just given by $A_\rms^2$. For the larger $w_\rms = \tw / 10$, we see how this factor is close to $A_\rms^2$ for small values, but go to $\m{A^2} > A_\rms^2$ as $\exp(-w_\rms^2\omega^2)$ goes to zero. Remarkably, the expression in \Eqref{eq:jitter-gauss} holds even for $w_\rms = \tw$, which has a significant probability of reordered arrival times.

\begin{figure}
\includegraphics[width=\textwidth]{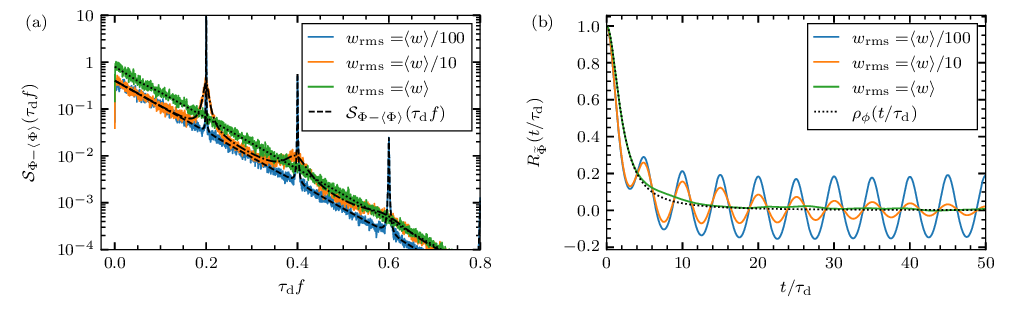}
\caption{Power spectral density (left) and autocorrelation function (right) for a sum of Lorentzian pulses with normal waiting times according to a renewal process with mean value $\tw = 5 \td$, and various values of $w_\rms$. Left, the analytical expression in \Eqref{eq:renewal-gauss} is presented by the black dashed line. Right, the black dotted line denotes the pulse autocorrelation.}\label{fig:renewal-gauss}
\end{figure}

For a renewal process with normally distributed waiting times, the polar form of the characteristic function gives $r = \exp\left( - w_\rms^2 \omega^2 / 2\right)$ and $\theta = \tw \omega$. Thus, in \Eqref{eq:PSD_fpp_renewal_polar} we get
\begin{equation}
   \psd{\Sn}{\omega} = \td \gamma A_\rms^2 I_2 \psdsnw(\td \omega) + \td\gamma \m{A}^2 I_2 \psdsnw(\td \omega) \frac{\sinh\left(w_\rms^2 \omega^2 /2 \right)}{\cosh\left(w_\rms^2 \omega^2 /2 \right) - \cos(\tw \omega)}, \quad \omega>0.
  \label{eq:renewal-gauss}
\end{equation}
Note that we have disregarded the $\delta(\omega)$-term and we have used $\m{A^2} = \mA^2+A_\rms^2$.
This power spectrum is presented in \Figref{fig:renewal-gauss}. For the smallest $w_\rms$-value, the PSD is quite similar to the one for the normally distributed jitter, although there is a broadening of the higher-frequency spectral peaks which is not seen in the jittered process. As $w_\rms$ increases in value, the peaks broaden further. For $w_\rms=\tw/10$, only the two first peaks are visible while for $w_\rms=\tw$, the PSD is quite similar to the pure exponential decay of the Lorentzian pulse function. This is also seen in the right-hand side of \Figref{fig:renewal-gauss}, where the black dashed line gives the autocorrelation function of the Lorentzian pulse.

Also note that the behavior of the monotonically decaying part of the power spectrum is the same as in \Figref{fig:jitter-gauss}: For $w_\rms = \tw/100$, the spectrum stays at the $A_\rms^2$-level for the entire plotted frequency range, while for $w_\rms = \tw$, it stays at the $\m{A^2}$-level. For the intermediate case, $w_\rms = \tw/10$, the level changes from $A_\rms^2$ to $\m{A^2}$ at around $0.3<\td f <0.6$.

We emphasize that although the calculations that lead up to \Eqsref{eq:jitter-gauss} and \eqref{eq:renewal-gauss} are only valid for a fixed order of the arrival times (which does not allow for negative waiting times or for jitters which exceed the periodicity), this is not evident from the power spectra in \Figsref{fig:jitter-gauss} and \ref{fig:renewal-gauss}: even for $w_\rms = \tw$, the analytic expressions agree with the realizations of the time series.

Very close to the periodic case, $w_\rms/\tw \ll 1$, we might expect the jitter and renewal processes to have the same behavior, but this is not the case. 
Approximating the maxima in the fraction  in \Eqref{eq:renewal-gauss} to $\tw \omega = 2 \pi k$  gives 
\begin{equation}
  \frac{\sinh\left[(2\pi w_\rms k/\tw)^2 /2 \right]}{\cosh\left[(2\pi w_\rms k/\tw)^2 /2 \right] - 1},
  \label{eq:renewal-gauss-maxima}
\end{equation}
for which the first terms in the Taylor expansion are 
\begin{equation}
  \frac{4}{(2\pi w_\rms k/\tw)^2} + \frac{(2\pi w_\rms k/\tw)^2}{12}.
  \label{eq:renewal-gauss-maxima_taylor}
\end{equation}
Thus the renewal process displays a power-law decay of the maxima that is not seen for the mass of the delta spikes in \eqref{eq:jitter-gauss}, which follow the characteristic function of the jitter process.

These examples show that semiperiodic phenomena in for example turbulent fluids cannot be expected to produce more than the first two peaks of the Dirac comb in the PSD, and that chaotic dynamics do not need to deviate far from strict periodicity in order for the Dirac comb to disappear. We will in \Secref{sec:rb-fpp} see an example of this behavior for the turbulent bursting in thermal convection. We further note that in practice, this effect may hinder attempts at estimating the spectral decay from periodic peaks in the PSD of a measurement time-series --- a very strict periodicity is required to get a good estimate even if the waiting time distribution is clearly unimodal.

\subsubsection{Comparison between different waiting time distributions}

Using $\beta=(\tw/w_\rms)^2$ for Gamma distributed waiting times and choosing $\beta$-values to correspond to the $w_\rms$-values in \Figref{fig:renewal-gauss} gives spectra which are visually indistinguishable from the results for the normally distributed waiting times, even for $w_\rms = \tw$ corresponding to $\beta =1$. This suggests that \Eqref{eq:renewal-gauss} is a generally useful approximation for many different waiting time distributions. Here, we compare the normal distribution to four other waiting time distributions. We only consider distributions with finite mean and variance, where the distribution is fully specified by $\tw$ and $w_\rms$. 

Gamma distributed waiting times are the natural generalization of the waiting times for a Poisson process, and are presented in \Secref{sec:gamma-wait}. For a distribution with power-law decay towards large values, we use inverse Gamma distributed waiting times with parameters $\alpha$ and $\beta$,
\begin{equation}
    p_w(w;\alpha,\beta) = \frac{\beta^\alpha w^{-\alpha-1}}{\Gamma(\alpha)} \exp(-\beta/w), \quad w>0.
\end{equation}
This PDF decays as $w^{-\alpha-1}$ for large $w$. The characteristic function is
\begin{equation}
    \psi_w(\omega) = \frac{2 (-\rmi \beta \omega)^{\alpha/2}}{\Gamma(\alpha)} \besselK_\alpha\left(\sqrt{-4 \rmi \beta \omega}\right), \quad \alpha = \frac{\tw^2}{w_\rms^2}+2, \quad \beta = \left( \frac{\tw^2}{w_\rms^2}+1 \right) \tw.
\end{equation}
Here, $\besselK$ is the modified Bessel function of the second kind \cite[10.25]{olverDLMFNISTDigital2025} and $\alpha>2$ ensures the variance of the waiting times is finite.
%
Both the Gamma and inverse Gamma distributions give strictly positive definite waiting times while the normal distribution does not. For small $w_\rms/\tw$, the latter is assumed to give a negligible effect, but for large $w_\rms/\tw$, the analytic result in \Eqref{eq:renewal-gauss} is no longer valid. 

\begin{figure}
\includegraphics[width=0.5\textwidth]{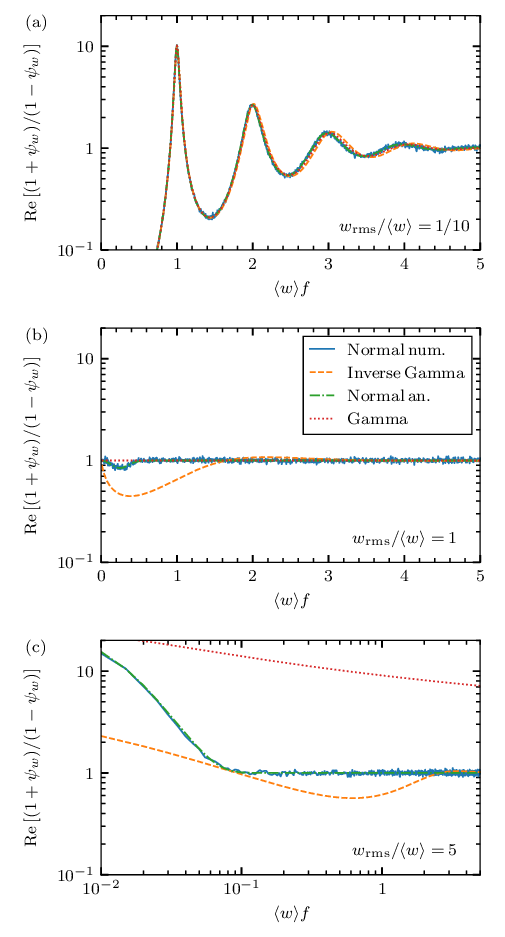}
\caption{The waiting time contribution to the power spectral density for various values of $w_\rms/\tw$. The legend in panel (b) is valid for all plot panels. 'an.' and 'num.' respectively refer to the analytic result, \Eqref{eq:renewal-gauss}, and the numerical result described in the text. Note the logarithmic frequency scale in panel (c).}\label{fig:psd-comparison-wait}
\end{figure}

In \Figref{fig:psd-comparison-wait}, we present results from analytic expressions for comparison between the normal, Gamma and inverse Gamma waiting time distributions. To isolate the effect of the waiting time distribution, we only show the $\Re[\cdot]$-part of \Eqref{eq:psd-snn-inf-renewal-arms}. Alternatively, the plots are for $A_\rms = 0$, $\td=\tw$ and pulse functions given by a Dirac delta.
In addition, we present the numerical result for the normal distribution, calculated as follows: first, we generate the arrival times by drawing normally distributed variables with $\tw=1$ and various $w_\rms$ and summing them up. Then the Fourier transform of the forcing in \Eqref{eq:S_split} is calculated directly as 
\begin{equation}
    \fourT{f}{\omega} = \tw \sum_{k=1}^{K(T)} \exp(-\rmi \omega s_k)
\end{equation}
and the PSD for this realization is estimated  as $\abs{\fourT{f}{\omega}}^2/T$. Finally, this process is repeated $M$ times to create an ensemble average for the PSD of the forcing. We set $T = 10^4 \tw$ and average over an ensemble of $M=10^3$ realizations.
In \subFigref{fig:psd-comparison-wait}{a}, we show that for close to periodic arrival times, $w_\rms/\tw = 1/10$, the actual waiting time distribution matters little for the spectrum and the normal distribution is a good approximation. In \subFigref{fig:psd-comparison-wait}{b}, the Gamma distribution reduces to the exponential distribution, giving a flat spectrum and the spectrum due to the normal distribution is very close to the one due to the Gamma distribution. The difference would be hard to distinguish for realizations of a process with a finite pulse duration and randomly distributed amplitudes. Here, the inverse Gamma distribution is markedly different from the Gamma and normal distributions due to its heavy tail. For highly variable waiting times compared to the mean waiting time, \subFigref{fig:psd-comparison-wait}{c}, the differences between the three distributions are evident. In particular, the case with Gamma distributed waiting times converges only algebraically towards unity with increasing frequency, while the inverse Gamma distribution gives a spectrum which decays to unity within $\tw f = 5$. The large probability of long waiting times given by the algebraic tail of the inverse Gamma distribution is of less consequence at high frequencies than the Gamma distribution, which has algebraic decay towards short waiting times.


For $w_\rms/\tw = 10^{-2}$ and $10^{-3}$, the Gamma and normally distributed cases are visually indistinguishable, and are therefore not presented. 
The Bessel function is very demanding to compute for $w_\rms/\tw \ll 1$, but as the power law tail of this distribution is only significant for larger $w_\rms/\tw$ we expect similar results for the inverse Gamma distribution as well, as it is already very close to the normal and Gamma distributions for $w_\rms/\tw=1/10$.

\subsubsection{Correlated waiting times}

\begin{figure}
\includegraphics[width=0.5\textwidth]{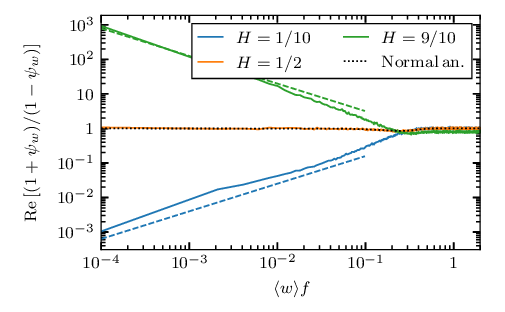}
\caption{The waiting time contribution to the power spectral density for various values of the Hurst parameter $H$ and $w_\rms/\tw=1$ for correlated pulse waiting times.  `an.' refers to the analytic result in \Eqref{eq:renewal-gauss}. The dashed lines give the low-frequency prediction for the PSD of the fractional Gaussian noise, $f^{1-2H}$ \cite{lowenFractalbasedPointProcesses2005}. }\label{fig:psd-comparison-wait-fGn}
\end{figure}

Here, we numerically investigate the effect of correlations between the waiting times.  We note that long-range correlations, algebraic decay and fractal properties of point processes is thoroughly investigated in \Ref{lowenFractalbasedPointProcesses2005}. To the best of the authors knowledge, the exact method presented here is new, as it allows for negative waiting times. To model correlations, we will use waiting times corresponding to the increments of the fractional Brownian motion: Letting $B_H(t)$ be the fractional Brownian motion with Hurst exponent $H$, we have
\begin{equation}
    w_k = \tw + w_\rms [B_H(k+1) - B_H(k)].
\end{equation}
The increments of $B_H$ are simulated using the Davis-Harte method, following the procedure in \Ref{diekerSPECTRALSIMULATIONFRACTIONAL2003}, and the results are presented in \Figref{fig:psd-comparison-wait-fGn}. For $H=1/2$, the waiting times are independent and follow the analytic prediction for normally distributed renewal waiting times as expected. For highly correlated $(H=9/10)$ and highly anti-correlated $(H=1/10)$ waiting times, the PSD of the stochastic process appears to converge to the expected PSD of the fractional Gaussian noise. These effects are only visible in the PSD of the stochastic model for low frequencies ($f<1/\tw)$, however, and the spectra are flat for higher frequencies. This is in contrast to semiperiodic waiting times, presented in \subFigref{fig:psd-comparison-wait}{a}, where the higher harmonics are visible for several multiples of $1/\tw$.

\subsubsection{Positive definite waiting times are not required.}
\begin{figure}
\includegraphics[width=0.5\textwidth]{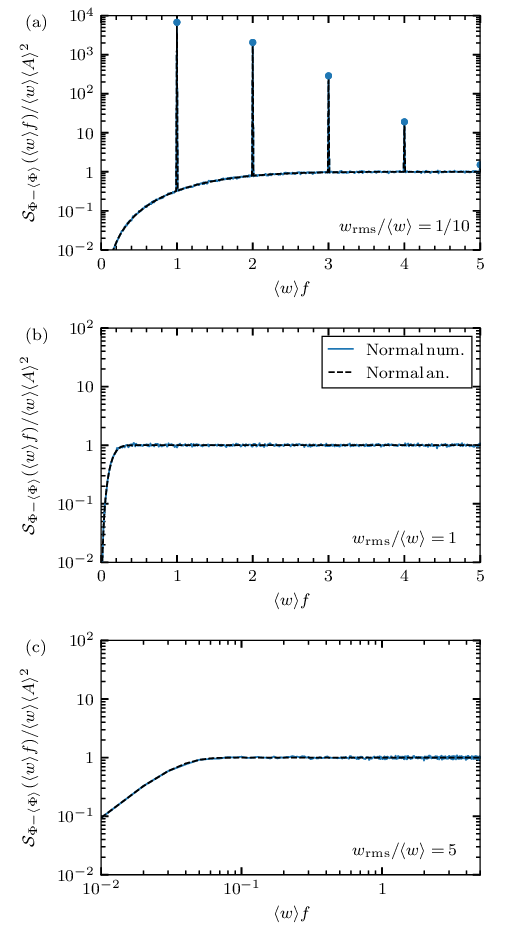}
\caption{The contribution to the power spectral density due to periodic arrivals with jitter for various values of $w_\rms/\tw$. The legend in (b) is valid for all plot panels. `an.' and `num.' refer respectively to the analytic result, \Eqref{eq:renewal-gauss}, and the numerical result described in the main text. Note the logarithmic frequency scale in panel (c).}\label{fig:psd-comparison-jitter}
\end{figure}

One striking feature of \Figref{fig:psd-comparison-wait} is the close correspondence between the numerical and analytical results for the normal distribution, despite the fact that normally distributed waiting times with $w_\rms \geq \tw$ give many negative waiting times. In \Figref{fig:psd-comparison-jitter} we demonstrate that the same holds for periodic arrival times with normally distributed jitter. These simulation results are obtained in the same way as for \Figref{fig:psd-comparison-wait}. This is despite the notion that jitter must be small to avoid reordering the arrival times \cite{beutlerRandomSamplingRandom1966}. In the same way, negative waiting times would give non-ordered arrival times.

Arrival times which do not maintain $s_{k+1}>s_k$ is, however, not necessarily an issue for the theory. The central requirements in \Secref{sec:fpp} are 1) wide sense stationarity of the process (for $T\to\infty$) to formulate the PSD, 2) the equality in distribution in \Eqref{eq:def-sk}, and 3) the relative narrowing of $p_K$ for large $T$. All are clearly fulfilled for periodic arrivals with jitter. For independent and identically distributed waiting times, all requirements should be fulfilled as long as the waiting times have a finite mean value and standard deviation. We discuss this issue further in \Appref{app:neg-wait}, indicating where potential problems may arise. For $\tw=0$, the process would be nonstationary, centered on the first waiting time. 

Practically, this means that if the waiting times of a process are estimated to have mean value $\tw$ and standard deviation $w_\rms$, simply assuming normally distributed waiting times should give a good estimate of the power spectral density of the process, as long as the waiting times are uncorrelated. This is demonstrated in the application to turbulent bursting in the thermal convection system in the next section. 

\section{Turbulent bursting as a sum of pulses}\label{sec:rb-fpp}

In this section, we apply the stochastic modelling framework to time series of the kinetic energy integral $\ke$ in numerical simulation of two-dimensional thermal convection. The convection model and its numerical implementation are detailed in \Appref{sec:rb}. The simulation data time-series of the kinetic energy integral are normalized by removing the mean value and dividing by the standard deviation
\begin{equation}
  \nke = \frac{\ke -\m{\ke}}{\ke_\rms} ,
  \label{eq:normalization}
\end{equation}
where the moments are estimated by the sample moments of the time series of $\ke$.

\begin{figure}
\includegraphics[width=\textwidth]{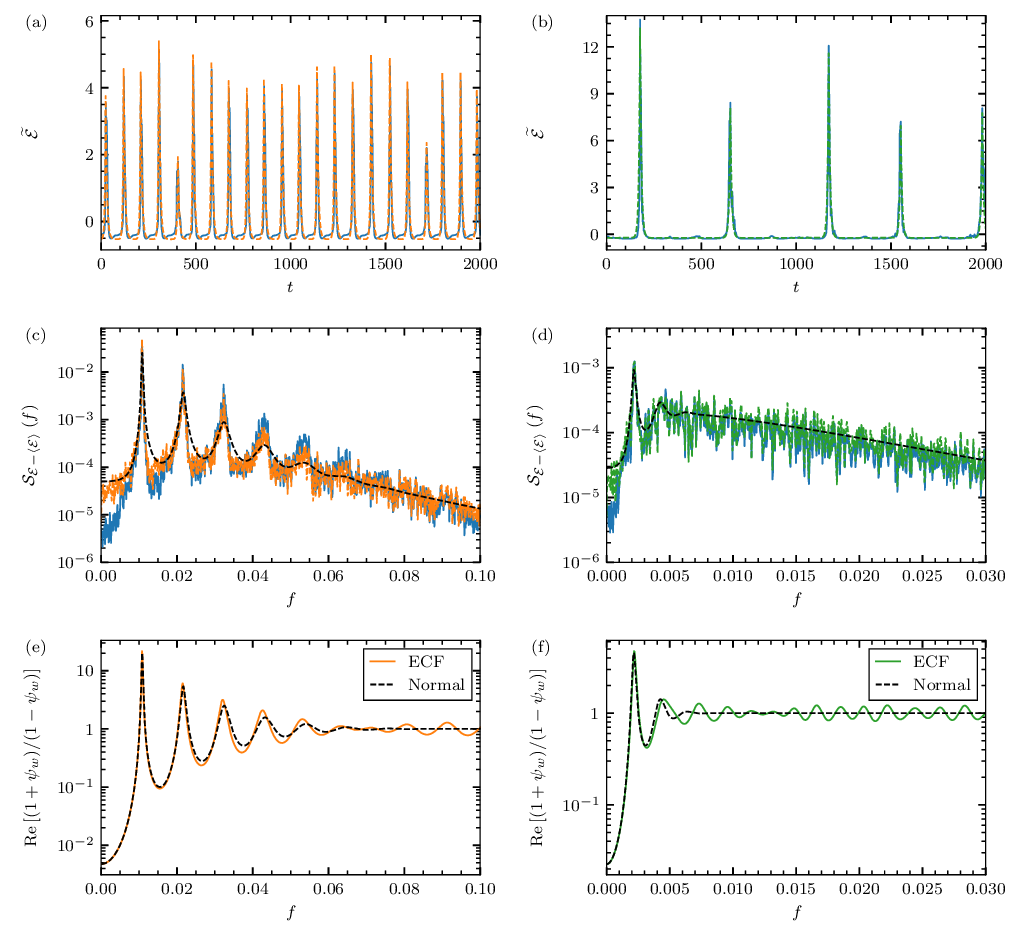}
  \caption{Time series and power spectral density of the kinetic energy of the fluctuating motions $\ke$ in turbulent thermal convection with $\mu = \kappa = 1.6\times10^{-3}$ (left) and $\mu = \kappa = 1.0\times10^{-4}$ (right). The data is fitted with a superposition of pulses shown in orange and green.}\label{fig:RB-TS-fit}
\end{figure}

\begin{figure}
\includegraphics[width=\linewidth]{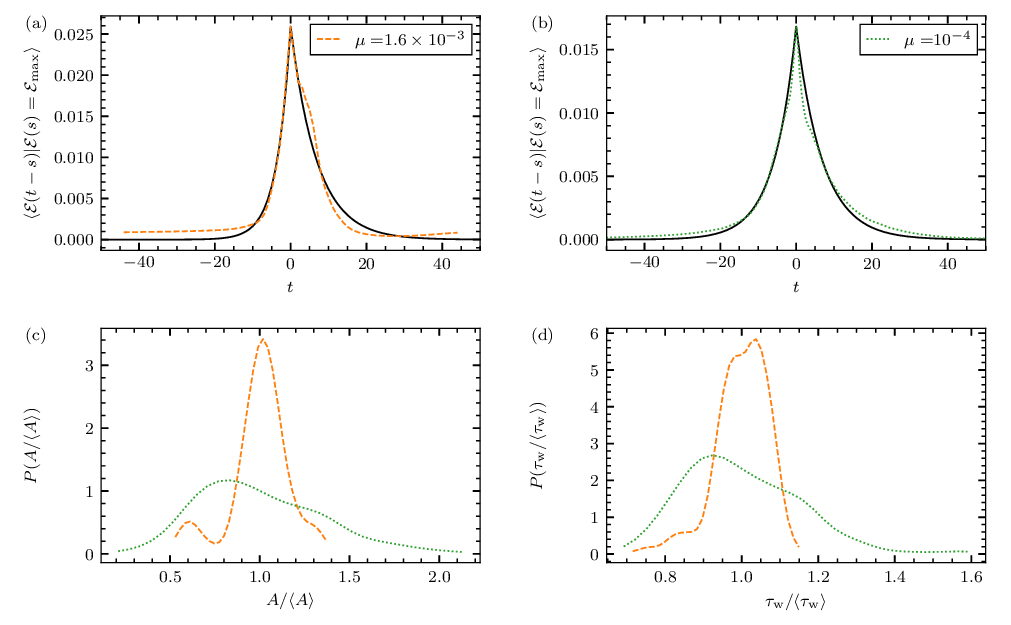}
  \caption{Estimated pulse properties of the kinetic energy of the fluctuating motions $\ke$ turbulent thermal convection. Top: Average burst shape and fitted exponential pulse for $\mu = \kappa = 1.6\times10^{-3}$ (left) and $\mu = \kappa = 1.0\times10^{-4}$ (right). Bottom: Amplitude and waiting time distributions of the pulses using Gaussian kernel density estimation.}\label{fig:RB-TS-cav}
\end{figure}

In \Figref{fig:RB-TS-fit}, the blue lines represent the fluctuating energy integral in the two cases $\mu = \kappa=1.6 \times 10^{-3}$ (left) and $\mu = \kappa=1.0 \times 10^{-4}$ (right). For the larger values of the heat diffusivity and viscosity, \subFigref{fig:RB-TS-fit}{a}, the turbulent bursts arrive frequently and close to periodically. This is consistent with the several clear peaks seen in the PSD in \subFigref{fig:RB-TS-fit}{c}. By contrast, the time between bursts appears more variable in the case of small heat diffusivity and viscosity coefficients, \subFigref{fig:RB-TS-fit}{b}, and accordingly only a single clear peak is present in the PSD in \subFigref{fig:RB-TS-fit}{d}. The orange and green lines are reproductions of the time series and power spectral densities using the stochastic model. The amplitudes and arrival times are found from the local maxima of the time series while demanding that each maximum is more than one pulse duration away from the rest. The pulse function is estimated by fitting a two-sided exponential pulse, defined in \Appref{app:exp_pulse}, to the average of all bursts in the time series as shown at the top of \Figref{fig:RB-TS-cav}. All stochastic model parameters are given in \Tabref{tab:RB-TS-fit}. 

From \subFigsref{fig:RB-TS-fit}{a}-(d), it is evident that the reproduced time series using the stochastic model captures most of the temporal and spectral features in the time series of $\ke$. The black dashed lines in \subFigsref{fig:RB-TS-fit}{c}-(f) show the normal waiting time approximation, using \Eqref{eq:renewal-gauss} with the values in \Tabref{tab:RB-TS-fit}. The normal distribution approximation captures most of the low-frequency behavior of the PSD as expected for $w_\rms/\tw < 1$. In \subFigsref{fig:RB-TS-fit}{e} and (f) we compare the normal distribution approximation to \Eqref{eq:psd-snn-inf-renewal-arms}, also using the values in \Tabref{tab:RB-TS-fit}. The characteristic function is estimated using the empirical characteristic function (ECF), which is simply the sample average over $\exp(2 \pi \rmi f w)$. 
The higher-frequency oscillations in the spectrum are better captured by the ECF estimate. In particular, note the oscillation in \subFigref{fig:RB-TS-fit}{e} with maximum at $f\approx 0.09$. This is present in the ECF fit as well, and so represents a feature of the PSD not captured by the normal approximation. 

The found amplitude- and waiting time distributions are presented in \subFigsref{fig:RB-TS-cav}{c} and (d). The case with larger heat diffusivity and viscosity (lower Rayleigh number) indeed has a narrower and therefore closer to periodic waiting time distribution, \subFigref{fig:RB-TS-cav}{c}. This is consistent with the many peaks visible in the spectrum, and agrees well with \Figref{fig:renewal-gauss}. The found $w_\rms$ is between the lowest and middle $w_\rms$-cases in \Figref{fig:renewal-gauss}, and so is between the cases with many visible spectral peaks and the case with only two broad peaks. The lower Rayleigh number case has a narrower amplitude distribution as well, see \subFigref{fig:RB-TS-cav}{d}. Both the amplitude- and waiting time distributions are consistent with the lower Rayleigh number case being closer to the threshold for periodic states.

\begin{table*}[t]
  \centering
  \begin{tabular}{c|c|c|c|c|c|c|c}
    $\kappa=\mu$ &$\tw$ &$w_\rms$ & $w_\rms/\tw$ & $\td$ & $\lambda$ & $\mA$ & $A_\rms$ \\
    \hline
    $1.6\times10^{-3}$ & $9.2\times 10$ & 6.4 & $6.9\times 10^{-2}$ & $1.1 \times 10$ & $3.4\times10^{-1}$ & $2.6\times 10^{-2}$ &$4.2 \times 10^{-3}$ \\
    $1.0\times10^{-4}$  & $4.6\times10^2$ & $6.9\times10$ & $1.5 \times 10^{-1}$ & $1.3\times10$ & $4.6\times10^{-1}$ & $1.7\times10^{-2}$ & $5.8\times10^{-3}$ \\
  \end{tabular}
  \caption{Parameters from the conditional averaging.}
  \label{tab:RB-TS-fit}
\end{table*}

\section{Conclusion}\label{sec:conclusion}

In this contribution, we have presented a stochastic model describing the power spectra of time series from non-linear dynamics as a superposition of pulses. The analytically investigated solutions comprise periodically arriving pulses, appropriate for non-linear oscillators, and pulses arriving according to a renewal process, which is more appropriate for chaotic dynamics. For the special case of Poisson arrivals, the spectrum of the arrivals is flat while for periodic arrivals the spectrum is a Dirac comb. In both cases, the spectral decay is fully determined by the spectrum of the pulse function which simply modulates the spectrum of the arrivals. Deviations from periodicity have been investigated in two ways, either as independently distributed waiting times (renewal arrivals) with a clear mode or as random perturbations to the original periodic arrivals (jittering).

For strictly periodic processes, the Dirac comb is a robust feature of the PSD, only removed by a vanishing mean of the pulse amplitudes. However, even for very modest deviations from strict periodicity, all harmonics except the lowest few are lost. This is true both for pulses randomly distributed around the periodic arrival time and for unimodal and narrow waiting time distributions. This demonstrates why spectra from even weakly nonlinear or chaotic systems do not display a Dirac comb in the PSD. Further, it is shown that in the case of small variations from strict periodicity, normally distributed waiting times gives a good approximation of the PSD for widely different underlying waiting time distributions.

Two mathematical aspects of the theory have been elucidated: 
First, we show how the Dirac delta contribution to the zero-frequency part of the spectrum may be rigorously found by keeping the observation time $T$ finite until the very end of the PSD derivation. Secondly, we show numerically that the expressions for renewal waiting times and jittered periodic arrivals hold even if negative waiting times are allowed (resulting in a significant reordering of the arrival times). We argue that the crucial assumption is a finite average waiting time, not positive definite waiting times. 

Finally, we demonstrate the applicability of the model and the normal approximation by estimating the power spectral density of energy fluctuations in turbulent two-dimensional thermal convection, using only the estimated pulses, amplitudes and arrival times from numerical simulations of the model.

The methodology developed here has numerous potential applications to time-series analysis. Here we have demonstrated the application to  numerical simulations of two-dimensional thermal convection, where identification of pulses was unambiguous. In real data, noise and pulse overlap may complicate the detection of pulses. With the methodology developed here, the frequency PSD may be used to estimate the average pulse duration and periodicity and the shape of the pulse function in the presence of non-Poisson arrival times and possibly discriminate between jitter and renewal waiting times. By isolating the effect of the waiting time distribution, the importance of other effects on the shape of the PSD such as correlations between amplitude, duration times, and waiting times may be assessed. The PSD method may be straightforwardly applied to quantify the change in the waiting time distribution with control parameters such as the Rayleigh number in the present example or to experimental measurement data with semiperiodic oscillations or bursting, such as sawtooth oscillations and edge localized modes in magnetically confined plasmas \cite{lazarusSawtoothOscillationsShaped2007,chapmanControllingSawtoothOscillations2011,zohmEdgeLocalizedModes1996,leonardEdgelocalizedmodesTokamaks2014}.

\section*{Acknowledgements}

This work was supported by the UiT Aurora Centre Program, UiT The Arctic University of Norway (2020) and the Troms{\o} Research Foundation under grant number 19\_SG\_AT. Discussions with M.~Rypdal and M.~Overholt are gratefully acknowledged.

A.~T.\ and O.~E.~G.\ contributed equally to and G.~D.\ supported conceptualization, methodology and writing - review and editing. A.~T.\ and G.~D.\ contributed equally to investigation, software and visualization. A.~T.\ and O.~E.~G.\ contributed equally to funding acquisition. A.~T.\ performed the formal analysis and writing - original draft.

\section*{Data availability statement}

The code generating the kinetic energy integral time series, the output time series data, the associated fit functions, and the code generating the figures are publicly available \cite{decristoforoUitcosmoDeviationsfromspectralDiraccombduetosemiperiodicpulsesPublication2025}.

\appendix

\section{The thermal convection model}\label{sec:rb}


As an example of semiperiodic oscillations in physical systems, \Secref{sec:rb-fpp}, we consider the model equations describing two-dimensional thermal convection \cite{finnNonlinearInteractionRayleigh1993,finnRoleSelfconsistentLagrangian1993,sugamaShearFlowGeneration1994,garciaConfinementBurstyTransport2003,garciaBurstingLargescaleIntermittency2003,garciaTwodimensionalConvectionInterchange2006}
\begin{subequations}
\begin{gather}
\left( \frac{\partial}{\partial t} + \mathbf{\widehat{z}}\times\nabla\psi\cdot\nabla \right) \Theta = \kappa\nabla^2\Theta , \label{temperature}
\\
\left( \frac{\partial}{\partial t} + \mathbf{\widehat{z}}\times\nabla\psi\cdot\nabla \right) \Omega + \frac{\partial\Theta}{\partial y} = \mu\nabla^2\Omega , \label{vorticity}
\end{gather}
\end{subequations}
where $\Theta$ describes the temperature deviation from hydrostatic equilibrium, $\psi$ is the stream function for the two-dimensional fluid velocity field $\mathbf{v}=\mathbf{\widehat{z}}\times\nabla\psi$, and $\Omega=\mathbf{\widehat{z}}\cdot\nabla\times\mathbf{v}=\nabla^2\psi$ is the associated fluid vorticity. The normalized heat diffusivity $\kappa$ and viscosity $\mu$ are related to the Rayleigh and Prandtl numbers by $R=1/\kappa\mu$ and $P=\mu/\kappa$, respectively.

All dependent variables are assumed to be periodic in the $y$-direction with periodicity length $L$, thus for the temperature we have $\Theta(y)=\Theta(y+L)$. In the $x$-direction the boundary conditions are taken to be stress-free,
\begin{subequations}
\begin{gather}
\psi(x=0) = \psi(x=1) = 0 ,
\\
\Omega(x=0) = \Omega(x=1) = 0 ,
\\
\Theta(x=0) = 1, \quad \Theta(x=1) = 0 .
\end{gather}
\end{subequations}

We define the zonal mean of the stream function as
\begin{equation}
    \psi_0(x,t) = \frac{1}{L}\int_0^L \text{d}y\,\psi ,
\end{equation}
and the associated mean flow is given by $v_0=\partial\psi_0/\partial x$. Accordingly, we define
the kinetic energy of the fluctuating motions as
\begin{equation}
  \ke(t) = \frac{1}{L}\int_0^1 \text{d}x \int_0^L \text{d}y\,\frac{1}{2}\,[\nabla(\psi-\psi_0)]^2 .
\end{equation}

For a sufficiently large Rayleigh number, this model displays semiperiodic oscillations of the convective energy integral $\ke$ \cite{sugamaShearFlowGeneration1994,garciaConfinementBurstyTransport2003,garciaBurstingLargescaleIntermittency2003,garciaTwodimensionalConvectionInterchange2006,groteRegularChaoticSpherical2000,groteDynamicsConvectionDynamos2001,christensenZonalFlowDriven2001,busseConvectiveFlowsRapidly2002,christensenZonalFlowDriven2002}.

Numerical simulations of this model have been performed in BOUT++, a C++ framework for writing plasma fluid simulations \cite{dudsonBOUTFrameworkParallel2009}. We choose a domain size $L=1$ and a resolution of $128 \times 128$ equidistant grid points.  A finite difference scheme is used in the $x$-direction and a spectral scheme in the $y$-direction. Time integration is performed with  the PVODE solver \cite{byrnePVODEODESolver1999}. All fields are initialized with random fluctuations. The transition phase from initial values to turbulence is not considered in the analysis presented in \Secref{sec:rb-fpp}.



\section{Fourier transform and power spectral density}\label{app:def_four}

The PSD of a random process $\Sn(t)$ on a domain of duration $T$ is defined as
\begin{equation}
  \psd{\Sn}{\omega} = \lim_{T \to \infty} \frac{1}{T} \m{\abs{\fourT{\Sn}{\omega}}^2},
\end{equation}
where the angular brackets denote an average over all random variables and
\begin{equation}
  \fourT{\Sn}{\omega} = \int\limits_{0}^{T} \rmd t\, \exp(-\rmi \omega t) \Sn_K(t)
\end{equation}
is the finite-time Fourier transform of the random variable over the domain $[0,T]$.

Analytical functions which fall sufficiently rapidly to zero (such as the pulse function $\snw$) have the Fourier transform 
\begin{equation}
  \four{\snw}{\vartheta} = \int\limits_{-\infty}^{\infty} \rmd \theta\,  \snw(\theta) \exp(-\rmi \theta \vartheta)
  \label{eq:fourier_trans}
\end{equation}
and the inverse transform
\begin{equation}
\snw(\theta) = \fourinv{\four{\snw}{\vartheta}}{\theta} =  \frac{1}{2 \pi} \int\limits_{-\infty}^{\infty} \rmd \vartheta \, \exp( \rmi \vartheta \theta ) \four{\snw}{\vartheta}.
\label{eq:inv_fourier_trans}
\end{equation}
Note that here, $\theta$ and $\vartheta$ are non-dimensional variables, as opposed to $t$ and $\omega$.

\section{Autocorrelation function for periodic arrivals}\label{app:ac-per}
In this appendix, we start from \Eqref{eq:PSD_fpp_finite_T_periodic} and derive the corresponding autocorrelation function. 
The inverse Fourier transform of the first term in the equation is
\begin{equation}
  \fourinv{\frac{1}{\tw} I_2 \m{ A^2 \tau^2 \psdsnw(\tau \omega)}}{t} = 
  \frac{1}{\tw} I_2 \m{ A^2 \fourinv{\tau^2 \psdsnw(\tau \omega)}{t}} =\frac{1}{\tw} I_2 \m{ A^2 \tau \acsnw(t/\tau)} 
  \label{eq:ac-per-1}
\end{equation}
The second term gives
\begin{align*}
  \fourinv{\frac{1}{\tw}\abs{\m{A \tau \fsnw(\tau \omega)}}^2}{t} &=  \frac{1}{\tw}\fourinv{\m{A \tau \fsnw(\tau \omega)}\m{A \tau \fsnw(\tau \omega)^*}}{t} \\
  &=  \frac{1}{\tw}\m{A_1 A_2 \tau_1 \tau_2\fourinv{\fsnw(\tau_1 \omega)\fsnw(\tau_2 \omega)^*}{t}} \\
  &=  \frac{1}{\tw}\m{A_1 A_2 \int\limits_{-\infty}^\infty \rmd s\, \snw\left(\frac{t+s}{\tau_1}\right) \snw\left(\frac{s}{\tau_2}\right)}
\end{align*}
while the third term gives 
\begin{multline*}
  \fourinv{\frac{2\pi}{\tw^2}\abs{\m{A \tau \fsnw(\tau \omega)}}^2\sum_{n=-\infty}^{\infty} \delta(\omega - 2 \pi n/\tw)}{t}  \\
  =\frac{2\pi}{\tw^2}\m{A_1 A_2 \tau_1 \tau_2 \fourinv{\fsnw(\tau_1 \omega)\fsnw(\tau_2 \omega)^*\sum_{n=-\infty}^{\infty} \delta(\omega - 2 \pi n/\tw)}{t}} \\
  =\frac{1}{\tw^2}\m{A_1 A_2 \tau_1 \tau_2 \sum_{n=-\infty}^{\infty} \fsnw(2 \pi \tau_1 n/\tw)\fsnw(2 \pi \tau_2 n/\tw)^* \exp(\rmi 2 \pi n t/\tw)} \\
  =\frac{1}{\tw}\m{A_1 A_2 \sum_{m=-\infty}^{\infty}\int\limits_{-\infty}^\infty \rmd s\, \snw\left(\frac{\tw m + t+s}{\tau_1}\right)\snw\left(\frac{s}{\tau_2}\right)}
\end{multline*}
where in the last line we used the Poisson summation formula \Eqref{eq:summation_3} with $a = 1/\tw$ and $s=t$, and the Fourier transform in the second term, above. In the end, we have that the autocorrelation is
\begin{multline}
  R_\Phi(t) = \frac{1}{\tw} I_2 \m{ A^2 \tau \acsnw(t/\tau)} - \frac{1}{\tw}\m{A_1 A_2 \int\limits_{-\infty}^\infty \rmd s\, \snw\left(\frac{t+s}{\tau_1}\right) \snw\left(\frac{s}{\tau_2}\right)} \\
  +\frac{1}{\tw}\m{A_1 A_2 \sum_{m=-\infty}^{\infty}\int\limits_{-\infty}^\infty \rmd s\, \snw\left(\frac{\tw m + t+s}{\tau_1}\right)\snw\left(\frac{s}{\tau_2}\right)}
  \label{eq:ac-per}
\end{multline}
In the case of degenerately distributed duration times, this expression simplifies to
\begin{equation}
  R_\Phi(t) = \gamma I_2 A_\rms^2 \acsnw(t/\td)  +\gamma\m{A}^2 I_2\sum_{m=-\infty}^{\infty}\acsnw\left(\frac{m}{\gamma} + \frac{t}{\td}\right)
  \label{eq:ac-per-deg-td}
\end{equation}

\subsection{The second moment of the periodic process}
Starting from \Eqref{eq:ac-per-deg-td} with $t=0$, we have that
\begin{align}
  \m{\Sn^2} &= R_\Sn(0) = \gamma I_2 A_\rms^2 + \gamma\m{A}^2 I_2\sum_{m=-\infty}^{\infty}\acsnw\left(\frac{m}{\gamma}\right) \label{eq:second-moment-per-deg-td}\\
  \rSn^2 &= R_\Sn(0) - \mSn^2 = \gamma I_2 A_\rms^2 + \gamma\m{A}^2 I_2\left[ \sum_{m=-\infty}^{\infty}\acsnw\left(\frac{m}{\gamma}\right) - \gamma \frac{I_1^2}{I_2}\right]
  \label{eq:rms-per-deg-td}
\end{align}
In the case of a highly intermittent process, $\gamma \ll 1$, only the $m=0$ term in the sum gives a contribution, $\acsnw(0) =1$, giving
\begin{equation}
\lim\limits_{\gamma \to 0} \frac{1}{\gamma}\,\Phi_{\rms}^2 = \langle{A^2}\rangle I_2,
\end{equation}
where we neglect the $\gamma^2$-contribution of the last term in the bracket in \Eqref{eq:rms-per-deg-td}. Thus, in the limit of no pulse overlap, the variance for the case of periodic pulses is equivalent to the case of Poisson arrivals, discussed in \Secref{sec:poisson}.

In the limit $\gamma \rightarrow \infty$, we can write $m/\gamma = m\, \rmd s \to s$ and treat the sum as an integral, $(1/\gamma)\sum_m \acsnw(m/\gamma)  \to \int \acsnw(s) \rmd s = I_1^2/I_2$, where the sum is over all integers and the integral is over all reals. The terms inside the bracket in \Eqref{eq:rms-per-deg-td} cancel, and we get
\begin{equation}
\lim\limits_{\gamma \to \infty} \frac{1}{\gamma}\,\Phi_{\rms}^2 = A_\rms^2 I_2.
\end{equation}
Since $A_\rms^2 = \langle{A^2}\rangle-\m{A}^2 \leq \langle{A^2}\rangle$, the periodic pulse overlap gives lower variance than the case of Poisson arrivals as there is less randomness in the process. For an exponential amplitude distribution, the variance in the periodic case is a factor two smaller. For amplitudes with zero mean value, it is equal to the Poisson case. For fixed amplitudes, the process has no variance as pulses will accumulate until the rate of accumulation exactly matches the rate of decay, after which the signal will remain constant.

\section{Sketch of argument for renewal process with negative waiting times}\label{app:neg-wait}

Here, we argue for the plausibility that negative waiting times will not change the PSD of the stochastic model driven by a renewal process, and indicate the weak points in the argument. We consider a process defined by the waiting times $w_k$ for $k \in \mathbb{Z}\setminus \{0\}$. All waiting times are independently and identically distributed with mean value $0 \leq \tw  \leq \infty$ and variance $0 \leq w_\rms^2\leq \infty$. We do not require $w>0$.

Let $s_0 = 0$ and define the positive and negative arrival times as 
\begin{equation}
    s_k = \sum_{l=1}^k w_l, \quad
    s_{-k} = -\sum_{l=-k}^{-1} w_l ,
\end{equation}
respectively. As may be verified by working out all cases, the definition of the sum of $k$ waiting times, \Eqref{eq:def-sk}, holds for all $n,k\in \mathbb{Z}$. Consequently, \Eqsref{eq:S0-convention}--\eqref{eq:S-convention} as well as $\psi_S(\sigma;k) = \psi_w(\sigma)^k$ also hold. Further, the $s_k$ will follow a central limit theorem: For large $k$ (positive or negative),
\begin{equation}
    s_k \sim \mathcal{N}(\tw k, w_\rms^2 \abs{k}).
\end{equation}
Note that the $s_k$ are no longer ordered due to the possibility of negative waiting times.

Further, let 
\begin{equation}
    I(T) = \{ k: 0<s_k<T \}
\end{equation}
be the set of all indices belonging to arrivals in the interval $(0,T)$ and define the counting process $K$ by $K(T) = \abs{I(T)}$, where $K(0)=0$ by definition. Then we may write the stochastic model on $(0,T)$ as
\begin{equation}
    \Sn(t) = \sum_{k\in I(T)} A_k \snw\left( \frac{t-s_k}{\tau_k} \right)
\end{equation}
and the full development up until \Eqref{eq:PSD_fpp_2} is equivalent if we replace $\sum_{k=1}^{K(T)}$ by $\sum_{k\in I(T)}$. The result is
\begin{multline}
  \psd{\Sn}{\omega} = \lim_{T\to\infty} \frac{\m{K(T)}}{T} I_2 \m{ A^2 \tau^2 \psdsnw(\tau \omega)}\\+ \abs{\m{A \tau \fsnw(\tau \omega)}}^2 \lim_{T \to \infty} \frac{1}{T} \sum_{K=0}^{\infty} P_K(K;T)\sum\limits_{\substack{k,l \in I(T)\\k\neq l}}\psi_S(\omega;l-k).
  \label{eq:PSD_fpp_neg_renewal}
\end{multline}
We note that this expression is independent of time $t$, indicating that the process $\Phi$ is stationary.

Now, we resort to looser arguments, in order of least to most questionable (in the authors opinion):

\begin{enumerate}
    \item It seems intuitively obvious that the possibility of negative waiting times will not change the mean value of $K$: for large $T$, $\m{K(T)} \approx T/\tw$. Thus, the first term in \Eqref{eq:PSD_fpp_neg_renewal} is equivalent to the first term in \Eqref{eq:PSD_fpp_2}.
    \item Due to the central limit theorem for $s_k$, it also seems reasonable that for large $T$, $P_K$ should still be a narrow distribution around $\m{K}$, allowing for replacing $K$ by $\m{K}$ in \Eqref{eq:PSD_fpp_neg_renewal}.
    \item There are end effects in the double sum over pulses in $I(T)$ in \Eqref{eq:PSD_fpp_neg_renewal}: we may for example have that $s_1 \notin I(T)$, but $s_{-2} \in I(T)$. Denoting the set of integers between $1$ and $K$ inclusive as $\kappa = \{1,\dots,K\}$, the double sum may be written as 
\begin{equation}
    \sum\limits_{\substack{k,l \in I(T)\\k\neq l}} = \sum\limits_{\substack{k,l \in \kappa \\k\neq l}}  - \sum\limits_{\substack{k,l \in \kappa \\ k,l \notin I(T)\\k\neq l}}  + \sum\limits_{\substack{k,l \notin \kappa \\ k,l \in I(T)\\k\neq l}} 
\end{equation}
The first term on the right gives the same expression as \Eqref{eq:PSD_fpp_2}. The second term on the right hand side remove all $s_1, \dots s_K$ which fall outside the interval $(0,T)$, while the last term on the right hand side adds all other arrival times which fall inside $(0,T)$. For large $T$ where the central limit theorem on arrival times holds, we may expect the vast majority of arrivals in $I(T)$ to also be in $\kappa$, suggesting that the last two terms on the right are negligible for $T\to \infty$. Consequently, the last term in \Eqref{eq:PSD_fpp_neg_renewal} is equivalent to the last term in \Eqref{eq:PSD_fpp_2}.
\end{enumerate}

If all the arguments above hold, the calculation from \Eqref{eq:PSD_fpp_neg_renewal} proceeds in exactly the same way as from \Eqref{eq:PSD_fpp_2}, ending in \Eqref{eq:PSD_fpp_renewal} which holds for waiting times which may take both positive and negative values. 

\subsection{Periodic arrivals with jitter}
As periodic arrivals follow a renewal process with degenerate waiting time distribution $p_w = \delta(w-\tw)$, all arguments above hold for purely periodic arrivals as well. Adding jitter adds some end effects, but far less than for non-degenerate renewal processes: the periodic arrivals with jitter have $s_k - \tw k \sim p_X$, so the end effects are constrained by the jitter rms-value, which is independent of $k$ or $T$. Therefore we expect the arguments above to hold very well for $T\to \infty$.

\section{Poisson summation formula}\label{app:Poisson_formula}

Here, we briefly present the well-known Poisson summation formula, which is treated in a number of textbooks, see Refs.~\onlinecite{bochnerLecturesFourierIntegrals1959,steinIntroductionFourierAnalysis1975,grafakosClassicalFourierAnalysis2014,overholtCourseAnalyticNumber2014}. For our purposes, the formulation used in Corollary VII.2.6 in Ref.~\onlinecite{steinIntroductionFourierAnalysis1975} is the most useful. The statement in the book is for functions on general Euclidian spaces, but we repeat it here only for our special case (the real line):

Suppose the Fourier transform of the function $h$ and its inverse are defined as in \Eqsref{eq:fourier_trans} and \eqref{eq:inv_fourier_trans}, respectively. Further suppose that $\abs{h(s)} \leq A(1+\abs{s})^{-1-\delta}$ and $\abs{\four{h}{\theta}} \leq A(1+\abs{\theta/2 \pi})^{-1-\delta}$ with $A>0$ and $\delta>0$. Then
\begin{equation}
    \label{eq:Poisson_summation_formula}
    \sum\limits_{m=-\infty}^\infty h(m) = \sum\limits_{n=-\infty}^\infty \four{h}{2 \pi n},
\end{equation}
where both series converge absolutely. Note that the inequality conditions guarantee that both $\abs{h(s)}$ and $\abs{\four{h}{\theta}}$ are integrable, which again guarantees that both $h$ and its Fourier transform are continuous and vanish at infinity (Theorem I.1.2 in Ref.~\onlinecite{steinIntroductionFourierAnalysis1975}).

Using properties of the Fourier transform, the summation formula can be cast to a number of different forms:
\begin{align}
\sum\limits_{n=-\infty}^{\infty} \four{h}{2 \pi n} &= \sum\limits_{m=-\infty}^{\infty} h(m) , \\
\sum\limits_{n=-\infty}^{\infty} a \four{h}{2 \pi n a} &= \sum\limits_{m=-\infty}^{\infty} h(m/a) , \\
\sum\limits_{n=-\infty}^{\infty} a \four{h}{2 \pi n a} \exp(i 2\pi n a s) &= \sum\limits_{m=-\infty}^{\infty} h(m/a + s).\label{eq:summation_3}
\end{align}
Here, $a$ is a positive constant.

The Poisson summation formula is used to compute the autocorrelation function from the PSD in \Appref{app:ac-per}.

\section{Pulse functions}\label{app:pulses}
Here, we present the pulse functions used in this contribution.

\subsection{Lorentzian pulse function}\label{app:lorentz_pulse}
The symmetric Lorentzian pulse is given by
\begin{equation} \label{Lorentzian}
    \snw(s) = \frac{1}{\pi(1+s^2)}.
\end{equation}
Its Fourier transform is
\begin{equation}
  \fsnw(\theta) = \exp(-\abs{\theta}),
\end{equation}
the integrals are \cite{garciaIntermittentFluctuationsDue2018} 
\begin{equation}
    I_n = \frac{\Gamma(n-1/2)}{\pi^{n-1/2} \Gamma(n)},
\end{equation}
and we have the pulse autocorrelation function
\begin{equation}
    \acsnw(s) = \frac{4}{4+s^2},
\end{equation}
and spectrum
\begin{equation}\label{eq:lonrentz_spectrum}
    \psdsnw(\theta) = 2 \pi \exp(-2\abs{\theta}).
\end{equation}

\subsection{Exponential pulse function}\label{app:exp_pulse}
The two-sided exponential pulse is given by \cite{garciaAutocorrelationFunctionFrequency2017}
\begin{equation} \label{eq:exp_pulse}
    \snw(s) = \begin{cases}
      \exp\left(\frac{s}{\lambda}\right), & s< 0,\\
      \exp\left(-\frac{s}{1-\lambda}\right), & s\geq0.
    \end{cases}
\end{equation}
Its Fourier transform is
\begin{equation}
  \fsnw(\theta) = \frac{1}{(\rmi + \lambda \theta)(\rmi -(1-\lambda)\theta)},
\end{equation}
the integrals are simply
\begin{equation}
  I_n = \frac{1}{n},
\end{equation}
and we have the pulse autocorrelation function
\begin{equation}
  \acsnw(s) = \frac{1}{1-2\lambda}\left[(1-\lambda)\exp\left(-\frac{\abs{s}}{1-\lambda}\right) - \lambda \exp\left(-\frac{\abs{s}}{\lambda}\right)\right],
\end{equation}
and spectrum
\begin{equation}\label{eq:exp_spectrum}
  \psdsnw(\theta) = \frac{2}{\left[1+(1-\lambda)^2\theta^2\right]\left[1+\lambda^2\theta^2\right]}.
\end{equation}

\section{Representation of the Dirac delta function}\label{app:delta-func}

A basic theorem in the theory of distributions is Theorem 2.5 in Ref.~\onlinecite{richardsTheoryDistributionsNontechnical1990}:
Let $h(x)$ be a piecewise continuous function such that:
\begin{itemize}
    \item $\int_{-\infty}^\infty \text{d} x\,\left| h(x) \right| < \infty$,
    \item$ \int_{-\infty}^\infty \text{d} x\,h(x) = 1$.
\end{itemize}
Writing $h_a(x) = a h(a x)$, we have that
\begin{equation}
    h_a(x) \to \delta(x) \quad \text{as} \quad a \to \infty.
\end{equation}
In particular, we note that
\begin{equation}
    h(x) = \frac{1-\cos(x)}{\pi x^2}
\end{equation}
fulfills the requirements of the theorem.

\section{Representation of delta functions under finite sampling}\label{app:finite_samples}

For graphical presentation of Dirac delta functions, we use its discrete analog, the Kronecker delta. For a time series with $N$ data points and time step $\dt$, the frequencies are $\omega_m = 2\pi m/(N \dt)$. A Dirac delta at a given angular frequency $\omega_*$ is then  given by $\delta(\omega_m - \omega_*) = \delta(2 \pi (m - k)/(N\dt)) = \frac{N \dt}{2\pi}\delta_{m-k}$, where $k$ is the nearest integer to $N\dt \omega_*/ 2\pi$. 




\section{The asymmetric Laplace distribution}\label{app:alap-dist}
The asymmetric Laplace distribution has several different parameterizations \cite{kozubowskiMultivariateAsymmetricGeneralization2000,reedNormalLaplaceDistributionIts2006,theodorsenProbabilityDistributionFunctions2018}. We choose a parametrization which has the exponential distribution as a straightforward limit \cite{theodorsenProbabilityDistributionFunctions2018}. Here, $\alpha>0$ is a scale parameter and $0<\lambda<1$ is an asymmetry parameter,
\begin{equation}\label{eq:alp-dist}
    p_X(x;\alpha, \lambda) = \frac{1}{2\alpha}\begin{cases}
    \exp\left( - \frac{A}{2\alpha(1-\lambda)}  \right), & A>0 , \\
    \exp\left( \frac{A}{2\alpha\lambda} \right), & A<0 . \\
    \end{cases}
\end{equation}
For this distribution, $\mA = 2\alpha (1-2\lambda)$ and $A_\rms^2 = 4 \alpha^2 (1-2\lambda + 2\lambda^2)$.

\bibliographystyle{apsrev4-2}
\bibliography{library.bib}

\end{document}